\documentclass[preprint2]{proto}
\usepackage{natbib, epsfig}
\bibpunct{(}{)}{;}{a}{,}{,}
\usepackage{times}

\usepackage[pdftex]{color}

\begin{document}








\title{\textbf{\LARGE Ages of Young Stars\footnotemark}}

\author {\textbf{\large David R. Soderblom}}
\affil{\small\em Space Telescope Science Institute, Baltimore MD USA}
\author {\textbf{\large Lynne A. Hillenbrand}}
\affil{\small\em Caltech, Pasadena CA USA}
\author {\textbf{\large Rob. D. Jeffries}}
\affil{\small\em Astrophysics Group, Keele University, Staffordshire, ST5 5BG, UK}
\author {\textbf{\large Eric E. Mamajek}}
\affil{\small\em  Department of Physics \& Astronomy, University of Rochester, Rochester NY, 14627-0171, USA}
\author {\textbf{\large Tim Naylor}}
\affil{\small\em School of Physics, University of Exeter, Stocker Road, Exeter EX4 4QL, UK}


\definecolor{green}{RGB}{0,160,50}
\newcommand{\david}[1] { \textcolor{cyan} {\ensuremath{\clubsuit} {\bf David:}  {#1}\ensuremath{\clubsuit} }}
\newcommand{\lynne}[1] { \textcolor{blue} {\ensuremath{\diamondsuit} {\bf Lynne:}  {#1}\ensuremath{\diamondsuit} }}
\newcommand{\rob}[1] { \textcolor{green} {\ensuremath{\heartsuit} {\bf Rob:}  {#1}\ensuremath{\heartsuit} }}
\newcommand{\eric}[1] { \textcolor{magenta} {\ensuremath{\spadesuit} {\bf Eric:}  {#1}\ensuremath{\spadesuit} }}
\newcommand{\tim}[1] { \textcolor{red} {\ensuremath{\spadesuit} {\bf Tim:}  {#1}\ensuremath{\spadesuit} }}

\begin{abstract}
\baselineskip = 11pt
\leftskip = 0.65in 
\rightskip = 0.65in
\parindent=1pc
{\small 

Determining the sequence of events in the formation of stars and planetary systems and their time-scales is essential for understanding those processes, yet establishing ages is fundamentally difficult because we lack direct indicators.  In this review we discuss the age challenge for young stars, specifically those less than $\sim$100 Myr old.  Most age determination methods that we discuss are primarily applicable to groups of stars but can be used to estimate the age of individual objects.  A reliable age scale is established above 20\,Myr from measurement of the Lithium Depletion Boundary (LDB) in young clusters, and consistency is shown between these ages and those from the upper main sequence and the main sequence turn-off -- if modest core convection and rotation is included in the models of higher-mass stars.    
Other available methods for age estimation include the kinematics of young groups, placing stars in Hertzsprung-Russell diagrams, pulsations and seismology, surface gravity measurement, rotation and activity, and lithium abundance.  We review each of these methods and present known strengths and weaknesses.  Below $\sim20$\,Myr, both model-dependent and observational uncertainties grow, the situation is confused by the possibility of age spreads, and no reliable absolute ages yet exist. The lack of absolute age calibration below 20\,Myr should be borne in mind when considering the lifetimes of protostellar phases and circumstellar material. \\~\\~\\~}
 
\end{abstract}  

\footnotetext{
Accepted for publication as a chapter in\\
Protostars and Planets VI\\
University of Arizona Press (2014)\\
eds. H. Beuther, R. Klessen, C.
Dullemond, Th. Henning.
}

\newpage

\cleardoublepage


\section{\textbf{
MOTIVATION AND SCOPE
}}

We know the timing of key events in the early history of our Solar System from meteorites because we can apply radiometric dating techniques to samples in a laboratory.  From this we infer the Sun's age, a very precise one, but even the Sun itself does not directly reveal its age in its visible properties.  The techniques for age-dating stars, and thereby other planetary systems, are of much lower precision. Establishing accurate or even precise ages for stars remains difficult, especially for young stars at the very stages where many interesting things, including planet formation and subsequent dynamical evolution,  are happening.

\noindent{\bf What does young mean?}
For the purposes of this review we consider ``young'' to be anything with an age ($\tau$) below about 100 Myr, which is to say the age of the Pleiades and younger.  For a coeval sample of stars, the higher stellar masses at this age will have begun core hydrogen burning and the highest masses will even have exhausted theirs. The lower stellar masses will be still in the pre-main sequence (PMS) phase of radial contraction which can take $\sim500$ Myr for objects just above the brown dwarf limit. 

In the context of understanding protostars and planets, we attempt here to establish time-scales and ordering of events {\em independently of the phenomena being studied}.  So, for instance, we note the presence of circumstellar material as suggestive of youth, but what we really want to do is to study the time evolution of circumstellar material and to bring to bear all applicable independent information.  We  cannot yet fully succeed in this goal, but progress is being made.

\noindent{\bf Goals in considering age estimation methods.} We wish, in order of increasing difficulty, to:
(1) Broadly {\em categorize} objects into wide age bins such as ``very young" or less than 5 Myr old stars generally associated with star-forming regions;  ``young pre-main sequence" stars which for solar mass are generally less than 30 Myr old; and ``young field" stars, generally less than a few hundred Myr old.
(2) Ascertain the correct {\em ordering} of phenomena, events, and ages.  
(3) Reliably detect {\em differences} in age, either for individual stars or for groups.
(4) Derive {\em absolute} ages, which requires not only sound methodology but also a definition of $\tau_0$, the evolutionary point at which age starts.

\noindent{\bf Recent reviews:} 
Ages and time-scales have always been implicit in the study of stars, and in recent years there have been several discussions of this subject.  
(1) \citet{soderblom:2010} wrote a general review of stellar ages and methods that emphasizes main sequence stars but includes a section on PMS objects.  Therein, age-dating methodologies are characterized as we do here into fundamental, semi-fundamental, model-dependent, and empirical categories.   
(2) The Soderblom review grew out of IAU Symposium 258 (``The Ages of Stars'') that was held in Baltimore in 2008.  The proceedings \citep{mamajek:2009} include reviews and contributions on PMS and young stars; see particularly \citet{jeffries:2009}, \citet{hillenbrand:2009}, and \citet{naylor:2009}.
(3) More recently \citet{jeffries:2012} has discussed age spreads in star-forming regions.
(4) \citet{preibisch:2012} has presented a thorough discussion of color magnitude diagrams (CMDs) and Hertzsprung-Russell diagrams (HRDs) for PMS stars and how ages are estimated from those diagrams.

There are also several chapters in the present volume that relate closely to our topic, particularly Bouvier et al., Dunham et al., and Alexander et al.


\section{
CRITICAL QUESTIONS ABOUT AGES}

\begin{enumerate}
\setlength{\itemsep}{-4pt}

\item
Can we establish reliable and consistent ages for young stars that are independent of the phenomena we wish to study? (Sec. \ref{sec:conclusions})

\item
What is the age {\em scale} for PMS and young stars and what inherent uncertainties affect it?  Can we construct a recommended age scale as a reference? (Sec. \ref{sec:ldb})

\item
We see scatter and dispersion in the HRDs and CMDs of PMS groups.  Is that evidence for an age spread?  Or can that be accounted for plausibly by unappreciated physics such as variable and differing accretion rates? (Sec. \ref{sec:agespread})

\item
Are age classifiers such as abundant lithium infallible or is it possible for a PMS star to have little or no lithium? (Sec. \ref{sec:li})

\item
Can the velocities of stars in groups reliably establish kinematic ages? (Sec. \ref{sec:kinematics})

\item
What is the chemical composition of young stellar populations, and how do uncertainties in composition affect stellar age estimation?  (Sec. \ref{sec:additional})

\item
What can be done to improve PMS ages over the next decade? (Sec. \ref{sec:conclusions})

\end{enumerate}

We will discuss individual methods and their precision and accuracy, as well as areas of applicability, weaknesses, and strengths.  We note that it is not strictly necessary for us to decide upon an absolute zero point to the age scale. This is a matter of both physical and philosophical debate (Sec.~\ref{sec:lumspread}), but all estimates of stellar age presently have uncertainties that exceed any uncertainty in the definition of this zero point.


\section{
SEMI-FUNDAMENTAL AGE TECHNIQUES
}

The only {\em fundamental} age in astrophysics is that of the Sun because the physics involved (decay of radioactive isotopes in meteorites) is completely understood and all necessary quantities can be measured.  {\em Semi-fundamental} methods require assumptions, but ones that appear to be well-founded and which are straightforward.
Because the models that predict the Lithium Depletion Boundary (LDB) at the very low mass end of cluster CMDs are physically simple, more so than for, say, cluster turn-offs, we recommend first establishing an age {\em scale} by considering clusters that have ages  from this technique.   Kinematic dating methods are also semi-fundamental in that they use a simple method with few assumptions, but, as we discuss, those assumptions may be invalid and the ages are problematic.


\subsection{
The Lithium Depletion Boundary (LDB)    
}
\label{sec:ldb}

\begin{table*}
\begin{tabular}{lcc@{\hspace{2mm}}ccc@{\hspace{2mm}}c@{\hspace{1mm}}c@{\hspace{2mm}}c}
\hline
\hline
Cluster & $I_{\rm LDB}$ &  LDB  & Ref. & $M_{\rm bol}$ & Homogeneous & Mermilliod MS & Overshoot MS & Ref. \\
        &    (mag)         & Age (Myr)    &        & (mag) & LDB Age (Myr) & Age (Myr) & Age (Myr) &  \\
\hline
$\beta$ Pic MG  &  &   $21\pm4$ &  a &  $8.28\pm0.54$ &  $20.3\pm3.4\pm1.7$ & &  & b \\
NGC 1960 &$18.95 \pm 0.30$ & $22\pm 4$ & c  & $8.57\pm0.33$& $23.2\pm 3.3 \pm 1.9$ & $<20$ & $26.3^{+3.2}_{-5.2}$ &  k\\
IC 4665  & $16.64 \pm 0.10$ & $28\pm5$ & d &$8.78\pm 0.34$ & $25.4 \pm 3.8 \pm 1.9$& $36\pm 5$ &  $41\pm 12$ &l  \\
NGC 2547 & $17.54\pm 0.14$ &$35 \pm 3$  & e & $9.58\pm 0.20$ & $35.4\pm 3.3 \pm 2.2$ & &  $48^{+14}_{-21}$ & m \\
IC 2602 & $15.64\pm 0.08$ & $46^{+6}_{-5}$  & f &$9.88\pm 0.17$ & $40.0 \pm 3.7 \pm 2.5$ & $36\pm 5$ & $44^{+18}_{-16}$ & m\\ 
IC 2391 & $16.21\pm 0.07$ & $50\pm 5$  & g &$10.31\pm 0.16$ & $48.6 \pm 4.3 \pm 3.0$ & $36\pm 5$  &$45 \pm 5$ & n \\
$\alpha$ Per & $17.70\pm 0.15$& $90\pm 10$  & h & $11.27 \pm 0.21$ & $80 \pm 11 \pm 4$ & $51\pm 7$ & 80 & o\\
Pleiades & $17.86 \pm 0.10$& $125\pm 8$  & i & $12.01 \pm 0.16$ & $126 \pm 16 \pm 4$ & $78\pm 9$ & 120 & o \\
Blanco 1 & $18.78\pm 0.24$& $132 \pm 24$  & j & $12.01 \pm 0.29$ & $126 \pm 23 \pm 4$ & & $115 \pm 16$ & j \\
\hline
\hline
\end{tabular}
\caption{LDB ages compared with ages determined from upper main sequence fitting using models both with and without convective overshoot. Columns 2-4 list the apparent $I$ magnitude of the LDB, the published LDB age and the source paper. Columns 5 and 6 give a bolometric magnitude and LDB age that have been homogeneously reevaluated using the locations of the LDB from the original papers, the evolutionary models of \citet{chabrier:1997} and bolometric corrections used in \citet{jeffries:2005}. The error estimates include uncertainty in the LDB location, distance modulus, a calibration error of 0.1 mag and then separately, a physical absolute uncertainty estimated from \citet{burke:2004}.
The last three columns give an upper main sequence age from \citet{mermilliod:1981} using models with no convective overshoot, followed by literature age estimate using models with moderate convective overshoot. References: (a) \citet{binks2013}; $I$ mag. not available; $M_{\rm bol}$ calculated from $K_{\rm LDB}$. (b) most massive member is A6V, hence no UMS age. (c) \citet{jeffries:2013}, (d) \citet{manzi:2008}, (e) \citet{jeffries:2005},  (f) \citet{dobbie:2010}, (g) \citet{barrado:2004}, (h) \citet{stauffer:1999}, (i) \citet{stauffer:1998}, (j) \citet{cargile:2010}, (k) \citet{bell:2013},  (l) \citet{cargile:2010b},  (m) \citet{naylor:2009}, (n) derived by E. Mamajek using data from \citet{hauck:1998} and isochrones from \citet{bertelli:2009}, (o) \citet{ventura:1998}.  
}
\label{ldbtab}
\end{table*}

\begin{figure}
\includegraphics[width=3.2in]{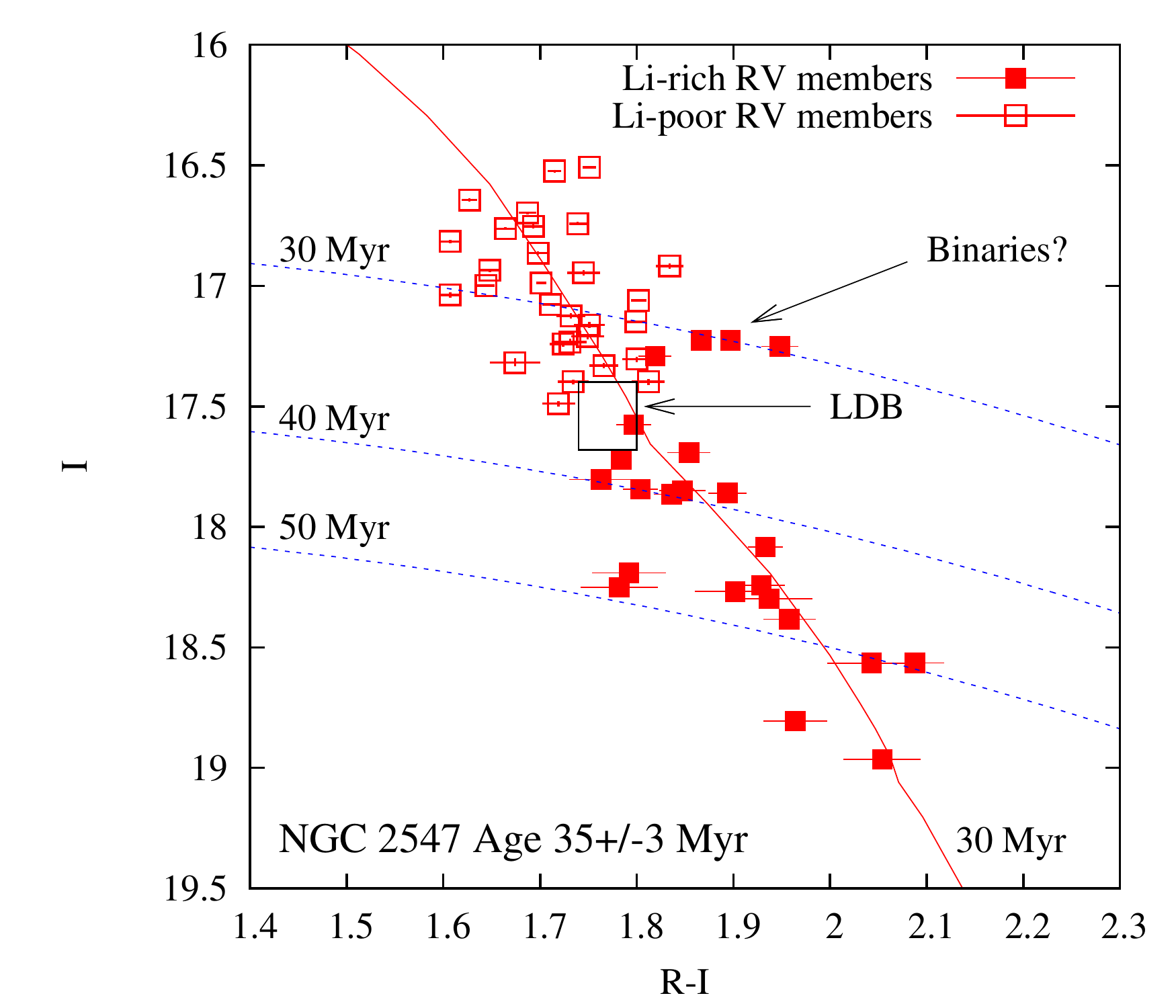}
\caption{Locating the LDB in the color-magnitude diagram of NGC 2547 \citep[adapted from][]{jeffries:2005}. The plot shows three dashed loci of constant luminosity, corresponding to LDB ages of 30, 40 and 50 Myr. The solid locus is a low-mass isochrone at 30\,Myr. The points represent cluster members that are found to possess a strong (undepleted) Li\,{\sc i} 6708\AA\ line or not and there is a reasonably sharp transition between these categories. A box marks the likely location of the LDB in this cluster and yields an age of $35.4 \pm 3.3$\,Myr (see Table~\ref{ldbtab}), but its precise position is made harder to judge by the presence of probable unresolved binary systems that appear over-luminous for their color. 
}
\label{fig:ldb_loc}
\end{figure}

As PMS stars age and contract towards the ZAMS, their core temperatures rise. If the star is more massive than $\simeq 0.06\,M_{\odot}$, the core temperature will eventually become high enough ($\sim 3$ MK) to burn lithium ($^{7}$Li) in 
proton capture reactions \citep{basri:1996, chabrier:1996, bildsten:1997, ushomirsky:1998}. PMS stars reach this Li-burning temperature on a mass-dependent timescale and, since the temperature dependence of the nuclear reactions is steep, and the mixing timescale in fully convective PMS stars is short, total Li depletion throughout the star occurs rapidly. This creates a sharp, age-dependent, lithium depletion boundary (LDB) between stars at low luminosities retaining all their initial Li and those at only slightly higher luminosities with total Li depletion.  

The LDB technique is ``semi-fundamental" because it relies on well-understood physics and is quite insensitive to variations in assumed opacities, metallicity, convective efficiency, equation of state and stellar rotation. \citet{burke:2004} conducted experiments with theoretical models, varying the input physics within plausible limits, and found that absolute LDB ages have ``1-$\sigma$" theoretical uncertainties ranging from 3\% at 200\,Myr to 8\% at 20\,Myr.  Comparisons of LDB ages predicted by a variety of published evolutionary models show agreement at the level of $\leq 10$\% across a similar age range. This model insensitivity applies only to ages determined from the LDB luminosity; the predicted $T_{\rm eff}$ at the LDB is much more sensitive to the treatment of convection and atmospheric physics. $L_{\rm bol}$ is also much easier to measure than $T_{\rm eff}$. Uncertainties in empirical bolometric corrections have a much smaller effect on derived ages than typical $T_{\rm eff}$ uncertainties from spectral types. In summary, the luminosity of the LDB is a good absolute age indicator, but the temperature of the LDB is {\it not} \citep{jeffries:2006}.

\begin{figure}
\includegraphics[width=3.2in]{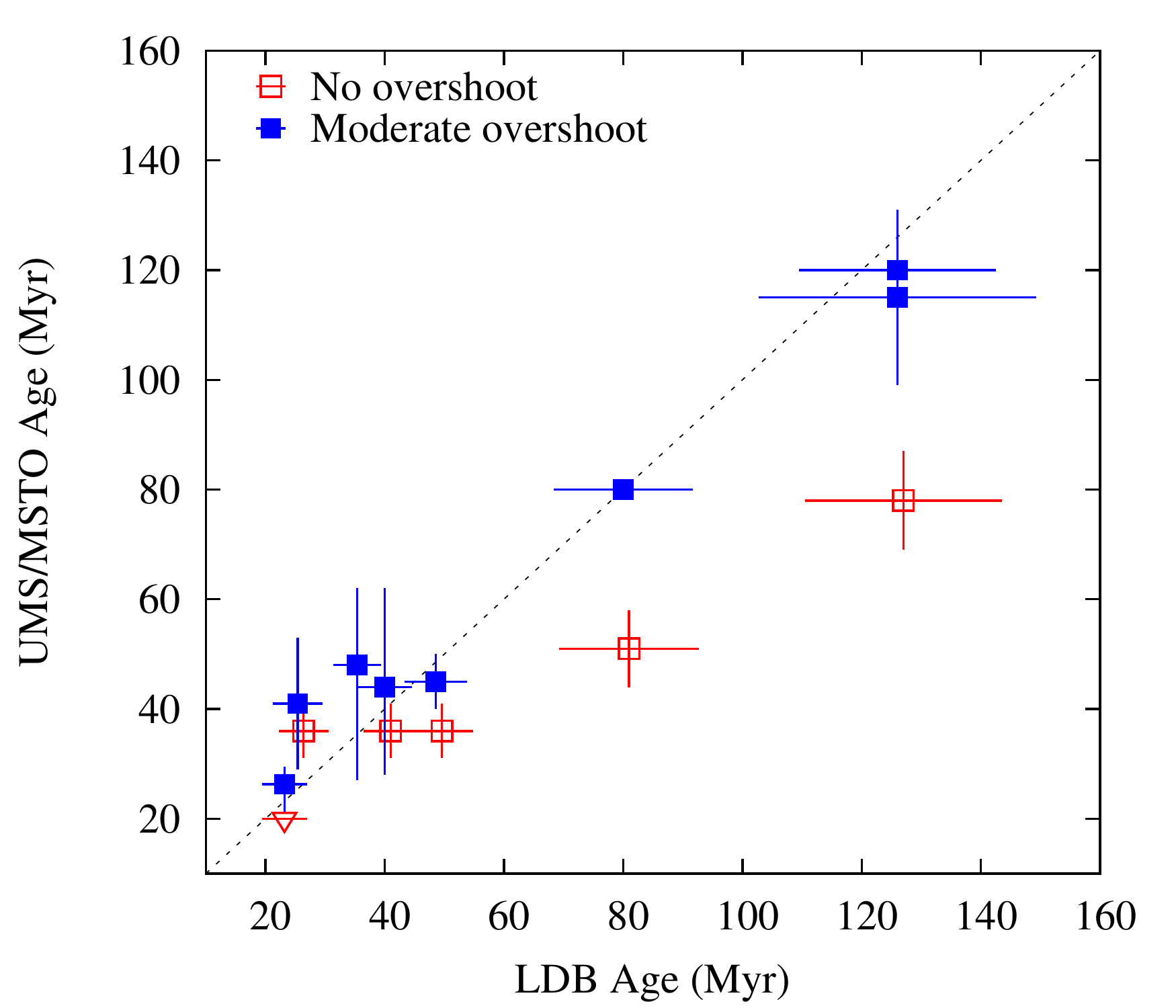}
\caption{A comparison of cluster LDB ages with ages determined from the upper main sequence (UMS) and main sequence turn-off (MSTO) using models without core convective overshoot and with a moderate amount of core overshoot (about 0.2 pressure scale heights). Data and sources are from Table~\ref{ldbtab}.}
\label{fig:ldb_vs_ums}
\end{figure}

LDB ages can be determined for coeval groups by determining the luminosity at which the transition from depleted to undepleted Li occurs; an example is shown in Fig.~\ref{fig:ldb_loc}.  This transition is expected to take $<5$\% of the stellar age, so the exact definition of ``Li-depleted'' is not crucial. This is useful since determining accurate Li abundances for cool stars ($T_{\rm eff} < 4000$\,K) is difficult. Instead one can rely on measuring the pseudo-equivalent width (EW) of the strong Li\, {\sc i}~6708\AA\ resonance feature. In an undepleted cool star EW(Li)\,$\simeq 0.6$\AA, falling to $<0.25$\AA\ in a similar star that has burned $>99$ per cent of its Li \citep{palla:2007}. 

LDB ages have been measured for nine clusters between 21 and 132\,Myr (see Table~\ref{ldbtab}); the published uncertainties are combinations of theoretical uncertainties and observational uncertainties associated with estimating the luminosity of the LDB. Contributing to the latter are distance estimates, reddening and bolometric corrections (for stellar luminosity not in the band observed), but these are usually small compared with the difficulty in locating the LDB in sparse datasets and where the low-mass stars may be variable and may be in unresolved binary systems. In younger clusters, any age spread of more than a few Myr might also blur the otherwise sharp LDB (see sec.~\ref{sec:agespread}). 

In Table~\ref{ldbtab} we homogeneously reevaluate these LDB ages using the LDB locations (and uncertainty), cluster distances and reddening given in the original LDB papers, but combined uniformly with the bolometric corrections used in \citet{jeffries:2005} and the evolutionary models of \citet{chabrier:1997}. Age uncertainties include those due to LDB location, an assumed 0.1 mag systematic uncertainty in both color and magnitude calibrations (usually negligible) and, presented separately, the systematic absolute age uncertainty estimated by \citet{burke:2004}. These results show that LDB age uncertainties have yet to attain the floor set by levels of theoretical understanding. There is scope for improvement, particularly in defining the position of the LDB, photometric calibration of cool stars and better estimates of cluster distances.

The LDB technique is limited by both physical and practical constraints to clusters of age 20-200\,Myr. Measuring the strength of the Li\,{\sc i}~6708\,\AA\ feature at sufficient spectral resolution ($R\geq 3000$) and signal-to-noise ratio to distinguish Li-depletion in very faint, cool stars is challenging. Table~\ref{ldbtab} lists the apparent Cousins $I$ magnitude of the LDB in the clusters where this has been possible. A photometric survey for candidates is required and subsequent spectroscopy must be capable of distinguishing cluster members from non-members or risk confusing Li-depleted members with older, unrelated field stars. Although the relationship between LDB luminosity and age does become shallower at older ages, leading to an increase in (fractional) age uncertainty, it is the faintness of the LDB that leads to the upper age limit on its applicability. At $\simeq 200$\,Myr, the LDB is at $L_{\rm bol}/L_{\odot} \simeq 5\times 10^{-4}$. The nearest clusters with ages $\geq 200$\,Myr are at distances of $\sim 300$\,pc or more, so finding the LDB would entail good intermediate resolution spectroscopy of groups of stars with $I>20$, beyond the realistic grasp of current 8-10-m telescopes. A lower limit to the validity of the technique is $\simeq 20$\,Myr and arises because at 10\,Myr some models do not predict significant Li destruction and at ages of 10-20\,Myr the difference in LDB-ages derived from different models is 20-30 per cent. 

Although possible in only a few clusters, the LDB method is well-placed to calibrate other age estimation techniques used in those same clusters. Its usefulness in estimating ages for individual stars is limited; the detection of (undepleted) Li in a low-mass star at a known distance and luminosity, or less accurately with an estimated $T_{\rm eff}$, will give an upper limit to its age. Conversely, the lack of Li in a similar star places a lower limit to its age. This can be useful for finding low-mass members of nearby, young moving groups or estimating the ages of field L-dwarfs \citep{martin:1999}, though might be confused by a weakening of the 6708\,\AA\ line in low gravity, very low mass objects \citep{kirkpatrick:2008}.

An example of how LDB ages can calibrate other techniques is provided by a comparison with ages determined for the same clusters from the main sequence turn-off (MSTO) and upper main sequence (UMS). These are derived by fitting photometric data with stellar evolutionary models, but depend on a number of uncertain physical ingredients including the amount of convective core overshooting and the stellar rotation rate \citep*[e.g.,][and see section~\ref{sec:cmd}]{maeder:1989, meynet:2000}. UMS/MSTO ages are listed, where available, in Table~\ref{ldbtab} using models with no convective core overshoot \citep{mermilliod:1981} and using models with moderate core overshoot (primarily the models of \citet{schaller:1992} with 0.2 pressure scale heights of overshoot). The comparison is shown in Fig.~\ref{fig:ldb_vs_ums}. As pointed out by \citet{stauffer:1998}, cluster ages determined from models with no core overshoot are about 30\%  younger than LDB ages but there is better agreement if moderate overshooting is included.  Recent models incorporating rotation for high mass stars show that this can mimic the effects of overshooting in extending main sequence lifetimes \citep{ekstrom:2012}. Hence LDB ages are consistent with evolutionary models that incorporate overshooting, rotation or a more moderate quantity of both, and comparison with LDB ages alone is unlikely to disentangle these. A homogeneous reanalysis of the UMS and MSTO ages for clusters with LDB ages, using uniform models and fitting techniques, would be valuable.

\smallskip
\noindent{\bf Summary of the LDB method:}
\begin{description}
\setlength{\itemsep}{-4pt}
\item[$+$] The LDB method involves few assumptions, and these appear to be on solid physical ground, making the method more reliable than others for clusters in the age range 20-200\,Myr.
\item[$+$] LBD ages depend only weakly on stellar composition and are insensitive to observational uncertainties.
\item[$+$] LDB observations require minimal analysis or interpretation: the Li feature is either clearly present or it is not.
\item[$+$] Age errors for this method appear to be $\sim$10-20\%, but could be lowered to $\sim$5\% with better observations.
\item[$-$] Detecting the presence or absence of Li at the LDB means acquiring spectra of moderate resolution for extremely faint objects, limiting its use to a few clusters.  It is unlikely that many more LDB ages will become available without larger telescopes.
\item[$-$] The age range for which the LDB method is applicable ($\sim$20-200 Myr) does not extend far past the ZAMS and does not cover star forming regions.
\item[$-$] The LDB method can only be applied to groups of stars of similar age, and provides only limits on the ages of individual low-mass stars.

\end{description}

\subsection{
Kinematic Ages
}
\label{sec:kinematics}

{\it Kinematic ages} theoretically can be derived from either 
proper motion or 3D velocity data for members of unbound, expanding associations by measuring either the group's expansion rate (``{\it expansion ages}") or estimating the epoch of smallest volume (``{\it traceback ages}"). Additionally, some authors have calculated ``{\it flyby ages}" by calculating the time of minimum separation between stellar groups, or individual stars and stellar groups, in the past. 
The 20th century work on kinematic ages is summarized in \citet{brown:1997}, while \citet{soderblom:2010} provides a review on more recent efforts.

Kinematic age techniques provide the hope of stellar age estimation independent of the many issues related to stellar evolution models; in particular they promise to yield absolute ages at $<20$\,Myr where the LDB technique is unavailable. However there are practical difficulties that have made the three common flavors of expansion ages unreliable. Historically, kinematic ages were calculated for OB subgroups, entire OB associations, and the entire Gould Belt complex. More recently, kinematic ages have been estimated for nearby young associations like the TW Hya and $\beta$ Pic groups \citep[e.g.,][]{delareza:2006,ortega:2002}. Kinematic techniques do not provide useful information on age spreads within star formation episodes, although runaways' ages have the potential to place lower limits on ages if ejected stars and birth sites can be unambiguously matched. 

{\subsubsubsection{OB Associations} 

\citet{brown:1997} discussed kinematic age estimates for OB subgroups based on proper motions alone. Table 1 of \citet{brown:1997} shows that rarely is there good agreement between either expansion ages or traceback ages versus ages derived via evolutionary tracks (``{\it nuclear ages}"). In most instances, kinematic ages were significantly shorter than nuclear ages. Through simple $N$-body simulations of unbound associations using realistic input parameters, \citet{brown:1997} 
demonstrated that calculating reliable kinematic ages of expanding associations, via either expansion rates or kinematic traceback, could not be estimated using proper motions alone because of their uncertainties; good radial velocities and parallaxes are needed.

{\subsubsubsection{Nearby Associations} 

The availability of the Hipparcos astrometric dataset provided the opportunity to attempt the estimation of kinematic ages using 3D velocities. The primary focus of these studies has been on newly discovered young associations within 100 pc \citep[e.g.,][]{zuckerman:2004}, with the primary targets being the TW Hya Association (TWA) and $\beta$ Pic Moving Group (BPMG). Some authors have elected to consider the kinematic ages of TWA and BPMG to be so well-determined and model-independent, that they have been used to judge the veracity of other age-dating techniques \citep[e.g.][]{song:2012}, however the veracity of the kinematic ages of TWA and BPMG are worth further scrutiny. We spend some time discussing both groups given the potential of their kinematic ages to act as benchmarks for other age-dating techniques. 

\subsubsubsection{TW Hya Association (TWA)} 

The most widely cited isochronal age for the TWA is that of \citet{webb:1999}: 10 Myr. \citet{makarov:2001} estimated an expansion age of 8.3 Myr, however the majority of stars used in the analysis were later shown spectroscopically to not be bona fide TWA members \citep{song:2003}. Later, a traceback age of 8.3\,$\pm$\,0.8 Myr for TWA was calculated by \citet{delareza:2006}, who simulated the past orbits of 4 TWA members using Hipparcos astrometry. However, as \citet{delareza:2006} admit, the traceback age for TWA hinged critically upon the membership of one system -- TWA 19 -- that turns out to be much more distant \citep[$d$ $\simeq$ 92\,$\pm$\,11 pc;][]{vanleeuwen:2007} than other TWA members \citep[$\bar{d}$ $\simeq$ 56 pc;][]{weinberger:2013}, and its position, proper motion, distance, and age are much more commensurate with membership in the $\sim$16 Myr-old Lower Centaurus-Crux association \citep{dezeeuw:1999, mamajek:2002, song:2012}. 

A subsequent calculation by \citet{makarov:2005} calculated an expansion age for TWA of 4.7\,$\pm$\,0.6 Myr by including two additional young stars (HD 139084 and HD 220476) in the analysis with three well-established members (TWA 1, 4, 11). However this age calculation appears to be unreliable as HD 220476 is a mid-G ZAMS (not pre-MS) star in terms of absolute magnitude and spectroscopic surface gravity \citep[e.g.][]{gray:2006}, and HD 139084 appears to be a non-controversial $\beta$ Pic group member \citep{zuckerman:2004}. 

Other attempts at calculating a kinematic age for TWA have been unsuccessful. Examining
trends of radial velocity versus distance, \citet{mamajek:2005} estimated a lower limit on the expansion age of TWA of 10 Myr (95\% confidence limit). The extensive trigonometric astrometric survey by \citet{weinberger:2012} was unable to discern any kinematic signature among the TWA members that might lead to a calculable kinematic age. It is odd that the first two studies which published well-defined kinematic ages both estimated the same age \citep[in agreement with the isochronal age published by][]{webb:1999}, and despite serious issues with regard to inclusion of non-members in both analyses. {\it We conclude that no reliable kinematic age has yet been determined for TWA}.

\subsubsubsection{$\beta$ Pic Moving Group (BPMG)} 

\citet{zuckerman:2001} estimated an isochronal age of 12$^{+8}_{-4}$ Myr for BPMG. Traceback ages of 11-12 Myr for BPMG have been calculated by \citet{ortega:2002, ortega:2004} and \citet{song:2003}. \citet{ortega:2002} simulated the orbits of the entire BPMG membership list from \citet{zuckerman:2001}, and showed that their positions were most concentrated 11.5 Myr ago (which they adopt as the age), and that at birth the BPMG members spanned a region 24 pc in size. \citet{ortega:2004} later revised their traceback analysis, and derived an age of 10.8\,$\pm$\,0.3 Myr for BPMG. \citet{song:2003} added new Hipparcos stars to the membership of BPMG, and found that ``{\it excepting a few outliers... all [BPMG] members were confined in a smaller space about $\sim$12 Myr ago}." \citet{song:2003} included $\sim$20 BPMG stars which appeared to be close together $\sim$12 Myr ago, but rejected half a dozen other systems which showed deviant motion. 

While these results seem self-consistent, they are inconsistent with the new LDB age of $21\pm4$ Myr determined by \citet{binks2013} and the question should be raised as to whether the inclusion or exclusion of individual members impacted the age derived from the traceback analysis. \citet{makarov:2007} found a wide distribution of flyby ages between individual $\beta$ Pic members and the group centroid, consistent with age 22\,$\pm$\,12 Myr. A new analysis of the kinematics of the BPMG by E. Mamajek (in preparation), using revised Hipparcos astrometry for the BPMG membership from the review by \citet{zuckerman:2004}, was unable to replicate the Ortega et al. and Song et al. kinematic age results. The expansion rate of the BPMG stars in $U$, $V$, $W$ velocity components versus Galactic coordinates results in an ill-defined expansion age of 25$^{+14}_{-7}$ Myr. More surprisingly, it was found that BPMG was not appreciably smaller $\sim$12 Myr ago, and fewer than 20\%\, of BPMG members had their nearest ``flyby" of the BPMG centroid during the interval 8 to 15 Myr ago. Statements about the size of the group reflected dispersions measured using 68\%\, intervals, hence they are reflecting the overall behavior of the stars, and hence not subject to rejection of individual members because they did not produce a desired outcome (past convergence). 

One is left with two interpretations: either a small subset of BPMG stars are kinematically convergent in the past (and the current membership lists are heavily contaminated with interlopers), or the current BPMG membership lists contain mostly bona fide members, but that the group is not convergent in the past as previously thought. Either way, {\it the kinematic ages for BPMG appear to be unreliable}. 

\subsubsubsection{Flyby Ages} 

Several studies have attempted to estimate ``flyby ages" of stellar groups by calculating when in the past
certain groups were most proximate, or when individual stars were closest to a group. The relation between
the flyby times and group ages is ambiguous. The turbulent nature of molecular clouds may give way to random motions
in star-forming complexes, and that means that tracing the bulk motions of various star-formation episodes back in time may lead
to minimum separation times which are unrelated to the epoch of star-formation. A few intriguing cases have appeared 
in the literature which deserve consideration. \citet{mamajek:2000} noted that the newly found $\eta$ Cha and $\epsilon$ Cha
groups were in close proximity in the past, and a Galactic orbit simulation by \citet{jilinski:2005} set their minimum separation 
at only a few pc 6.7 Myr ago -- similar to the isochronal ages of both groups. However, subsequent investigations have convincingly shown that the $\eta$ Cha and $\epsilon$ Cha groups have significantly different ages based on considerations of color-magnitude positions and surface gravity indices \citep{lawson:2009}, and a more recent kinematic analysis found that the groups were not much closer in the past than their current separation of $\sim$30 pc \citep{murphy:2013}. At present, {\it there appears to be no reliable
kinematic age for the $\epsilon$ Cha and $\eta$ Cha groups}. 

Another interesting flyby age was calculated by \citet{ortega:2007}, who demonstrated that the AB Doradus Moving Group (ABDMG) and the Pleiades were in close ($\sim 40$~pc) proximity 119\,$\pm$\,20 Myr ago, commensurate with modern ages for both groups \citep{stauffer:1998, barrado:2004, luhman:2005, barenfeld:2013}. Hence both could have formed in the same OB association \citep{luhman:2005}. However the uncertainties in the velocities of both ABDMG and the Pleiades are at the $\sim$0.5 km/s level, so the minimum uncertainty in their 3D positions $\sim$120 Myr ago is $\sim$60 pc per coordinate, and hence the uncertainty in their mutual separation is $\sim$100 pc. 

From consideration of the properties of OB associations as a whole (see upcoming discussion on ages of stellar associations), it is unclear whether one could ever reliably determine a kinematic age to better than $\sim$20 Myr accuracy for two subgroups (in this case ABDMG and Pleiades) that are assumed to have formed in the same OB association over scales of $\sim$10$^2$ pc. One would need to demonstrate that the groups formed in a very small volume, and one would need astrometry far superior than that delivered by Hipparcos to do so.
Subsequent work by \citet{barenfeld:2013} has shown that a non-negligible fraction of ABDMG ``stream" members are chemically heterogenous, and hence could have formed from multiple birth-sites unrelated to the AB Dor ``nucleus."  We conclude that the co-location of ABDMG and Pleiades $\sim$120 Myr ago is intriguing, but it is unclear whether the actual flyby age is sufficiently solid to reliably test other age techniques. 

\subsubsubsection{Runaway Ages}  

Runaway stars are those ejected from binaries after supernovae explosions or from the decay of higher multiple systems. In some cases their origins may be traceable to a particular star forming region or cluster, the traceback time giving their minimum age. The classic examples are AE Aur, $\mu$ Col and $\iota$~Ori, two individual O/B stars and an O/B binary which \citet{hoogerwerf:2001} suggested were ejected from the ONC 2.5\,Myr ago and hence that star formation was ongoing at least as long ago as this. However, $\iota$~Ori appears to be near the center of, and co-moving with its own, dense, $\sim$4-5 Myr-old cluster NGC 1980 \citep{alves:2012, pillitteri:2013}, perhaps casting doubt on the original location of the runaway stars and calling into question any conclusions regarding the ONC.  Without a population of runaway stars securely identified with a particular birth location it will be difficult to bring traceback ages to bear on the question of mean ages or age spreads. Lower mass runaways may be more plentiful, but will be harder to find and generally have smaller peculiar velocities \citep{poveda:2005,odell:2005}. Precise distances and proper motions from Gaia may open up this avenue of research and will clearly assist in the vital task of locating the origin of runaway stars.

\subsubsubsection{Inherent Uncertainties in Ages of Association Members}

There are inherent astrophysical uncertainties in the ages of members of stellar associations that can arise, particularly in the context of kinematics.  The assumption of single mean ages for members of large stellar associations is likely not a good one for groups larger in scale than typical embedded clusters and molecular cloud cores \citep{evans:2009}. Significant velocity substructure is detected within giant molecular clouds \citep{larson:1981}, and significant age differences are seen among subgroups of OB associations \citep{briceno:2007}. The combination of molecular cloud properties and observed properties of young stellar objects conspire to produce a characteristic timescale for the duration of star-formation $\tau_{\rm SF}$ over a region of length scale $\ell$:

\begin{equation}
\tau_{\rm SF} \sim \ell_{\rm pc}^{1/2} ~~~~{\rm (Myr)} 
\end{equation}

\noindent The relation comes from consideration of the observational data for star-formation over scales of 0.1 pc to 10$^{3}$ pc \citep{elmegreen:1996}, and the characteristic timescales for molecular clouds over similar scales \citep{larson:1981}. Hence the modeling of ``bursts" of star formation, and the implicit or explicit assumption of coevality of a stellar group should take into account empirical limits on the duration of star-formation in a molecular cloud complex over a certain length scale.  When adopting a {\it mean} age $\bar{\tau}$ for a {\it member} of an extended stellar association, one should naively predict a lower limit on the age precision for the star if the association's star-forming region was of size $\ell$. The fractional age precision 
for an individual group member when adopting a mean group age can be estimated as:

\begin{equation}
\epsilon = \frac{\delta \tau}{\bar{\tau}} \sim \frac{\tau_{SF}}{\tau} \sim \tau_{\rm Myr}^{-1} \ell_{\rm pc}^{1/2}
\end{equation}

\noindent For example, the nearest OB subgroup Lower Centaurus-Crux (LCC) has mean age $\sim$17 Myr and covers $\ell$ $\sim$ 50 pc in size \citep{mamajek:2002, pecaut:2012}, so we would naively predict a limit to the age precision of $\epsilon$ $\sim$ 50\% when adopting a mean group age for an individual member. Indeed, 
after taking into account the scatter in isochronal ages due to the effects of observational errors, the age spread in LCC has been inferred to be of order $\sim$10 Myr \citep{mamajek:2002,pecaut:2012}. 
It follows that adopting mean ages for members of entire associations which were larger than tens of pc in size becomes untenable (age errors $\epsilon$ $\sim$ 100\%) as one predicts $\delta \tau$ $\sim$ $\bar{\tau}$ -- unless one can convincingly demonstrate that the group was kinematically confined to a small region in the past. This is difficult to do due to substantial uncertainties in present day velocities, but may become possible by combining more precise Gaia astrometry with precise radial velocities.

\smallskip
\noindent{\bf Summary of kinematic methods:}
\begin{description}
\setlength{\itemsep}{-4pt}
\item[$+$] Methods are independent of stellar astrophysics.
\item[$+$] Gaia should provide precise astrometry for many faint members of young groups, and that will enable the determination of more statistically sound kinematic ages, and for more distant groups. 
\item[$-$] Kinematic ages derived using proper motions alone have been shown to be unreliable. Accurate radial velocities and parallaxes are required. 
\item[$-$] Recent traceback analyses appear to suffer from some degree of subjectivity in the inclusion or exclusion of individual group members (especially for the TWA and BPMG groups). 
\item[$-$] Some traceback ages have not held up when improved astrometric data comes available. 
\item[$-$] It is unclear whether any reliable, repeatable, kinematic mean age for a {\it group} has ever been determined. 
\item[$-$] Unless it can be shown that a kinematic group traces back to a very small volume there is no reason to suppose that these stars are coeval.
\end{description}


\section{
MODEL-DEPENDENT METHODS}

\subsection{
Placing Pre-Main Sequence stars in HRDs      
}
\label{sec:cmd}

This section discusses the methodology for comparing PMS evolutionary models with observations of young stars, including the effects of extinction/reddening, photometric variability, and ongoing accretion which make the task more challenging than comparable exercises for open and globular clusters.

\subsubsubsection{Models and Flavors of Models}

The theory of PMS evolution requires an appreciation of the physics that governs the radial gradients of density, pressure, temperature, and mass within stellar interiors, as PMS stars globally contract over time.  Our understanding of radiative transfer and consideration of the relevant energy sources (gravitational and light element nuclear burning) and opacity sources (atomic and molecular gas and possibly dust as well for the coolest stars), further leads us to predictions of observables.

Models such as those by \cite{dantona:1997,baraffe:1998,siess:2000,yi:2003,demarque:2004,dotter:2008,tognelli:2011} provide the radii, luminosities, and effective temperatures of stars of given mass at a given time.  These models may differ in their inputs regarding the equation of state, opacities, convection physics, outer boundary condition of the stellar interior, and treatment of atmospheres.  Additional physics such as fiducial initial conditions, accretion outbursts, ongoing accretion, rotation, and magnetic fields are also involved   \cite[e.g.,][]{palla:1999,baraffe:2009,baraffe:2012,tout:1999,hartmann:1998,baraffe:2002}.  From any of these evolutionary models, isochrones can be produced in either the natural plane of the theory ($L/L_\odot$ or g vs. $T_{\rm eff}$) or in any color-magnitude or color-color diagram used by observers.  As shown by e.g. \citet{hillenbrand:2008} the differences between ages predicted by the various theory groups increases towards younger ages and towards lower masses; there is $<$0.1-0.3 dex systematic variance in predicted ages at spectral type G2, but 0.25-0.6 dex at K6 among the models cited above.  See Figure~\ref{fig:hrd_age} for an example of isochrone differences.

\begin{figure}
\includegraphics[width=3.2in]{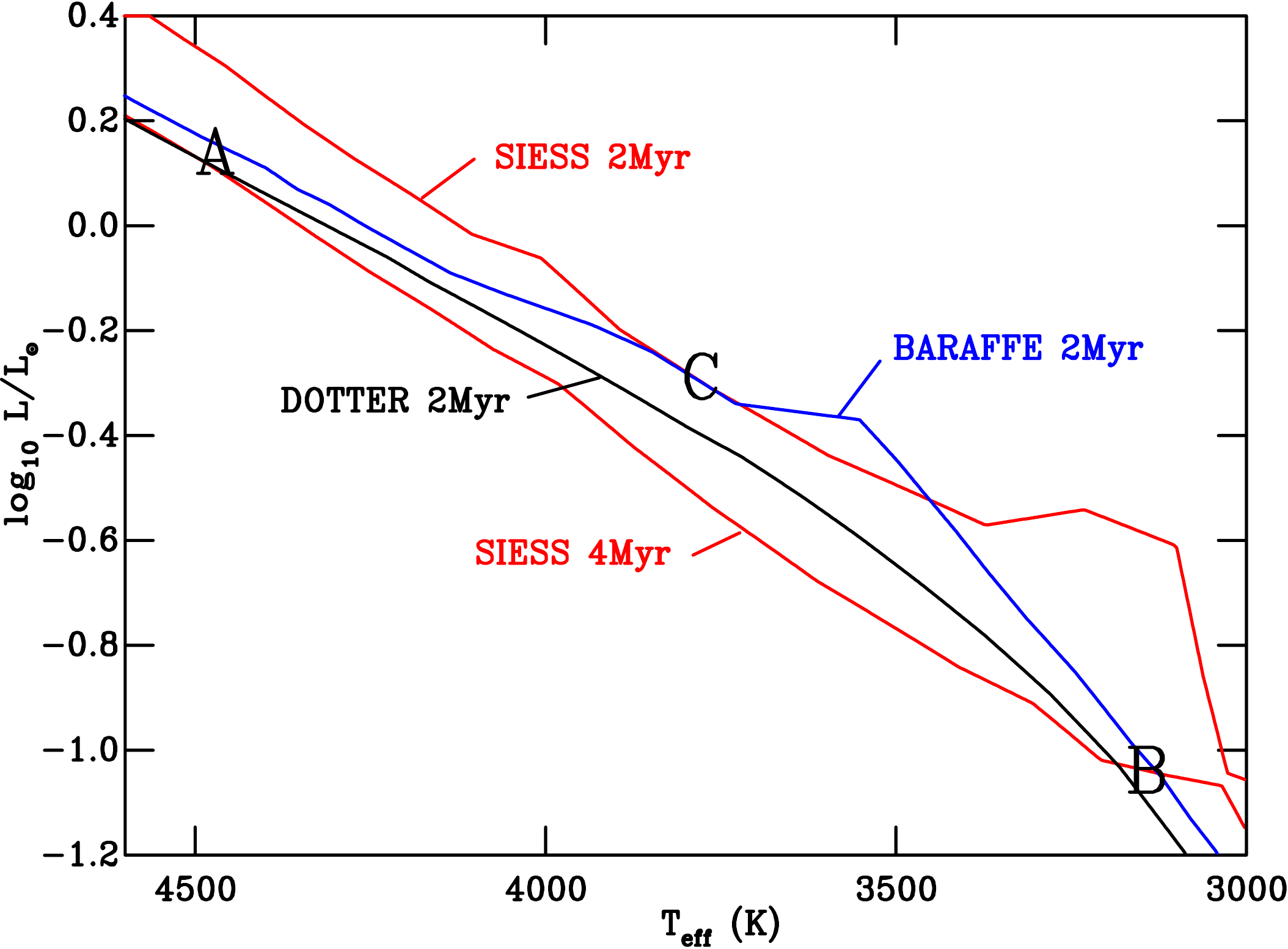}
\caption{An illustration of the difference in age obtained by using different interior models.
Stars at positions A or B would be given an age of 2 Myr using the interior models of \cite{dotter:2008} and \cite{baraffe:1998} but 4 Myr using the isochrones of \cite{siess:2000}.  A star at point C would be given an age of 2 Myr using the isochrones of \cite{siess:2000} and \cite{baraffe:1998} but 1.3 Myr on the isochrones of \cite{dotter:2008} (isochrone not shown).
}
\label{fig:hrd_age}
\end{figure}

\begin{figure}
\includegraphics[width=3.2in]{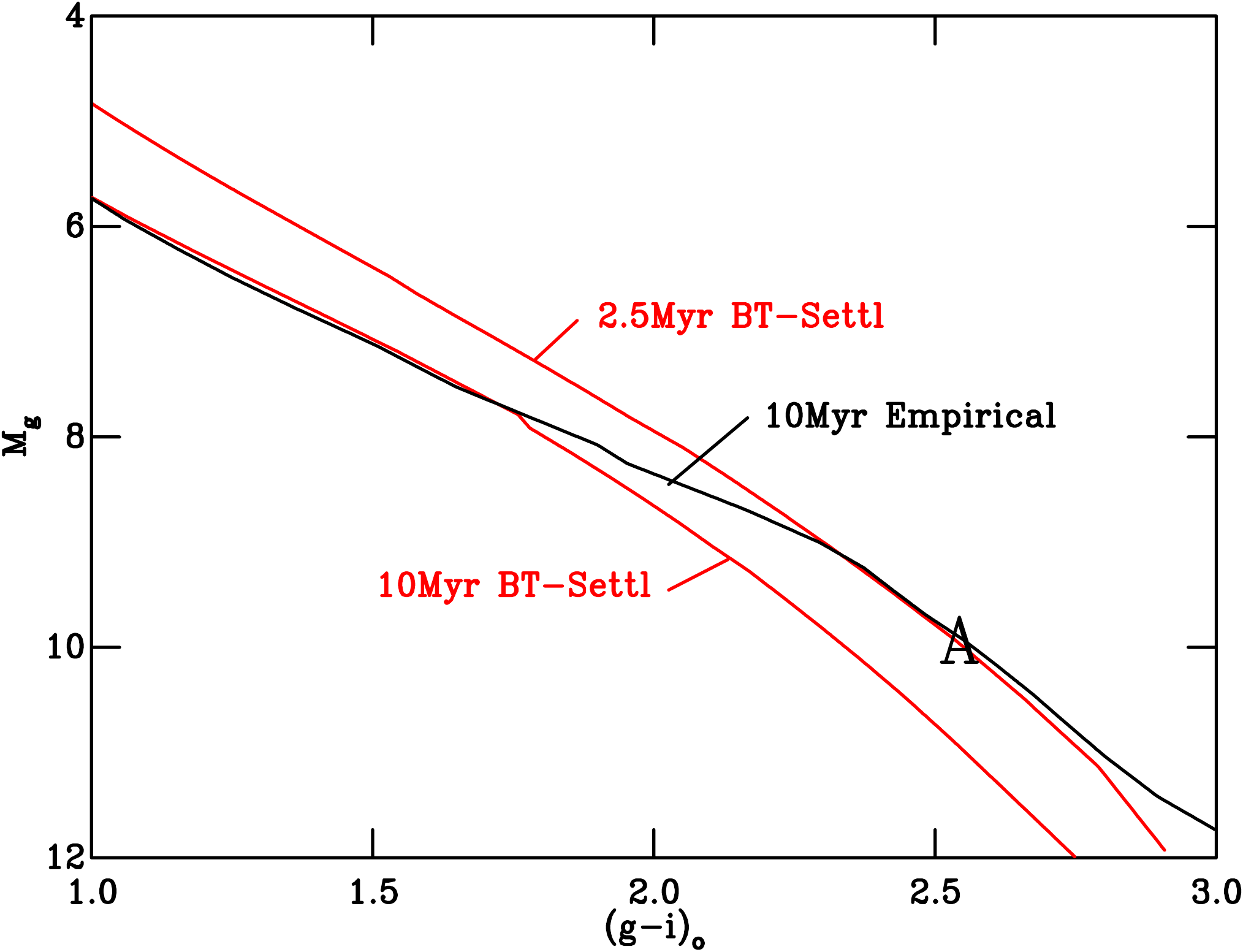}
\caption{An illustration of the difference in age obtained by using different color-effective temperature relationships.
A star at position A would be assigned an age of 2.5 Myr using the interior models of \cite{dotter:2008} and the BT-Settl model atmospheres of \cite{allard:2011}, but 10Myr using the same \cite{dotter:2008} interiors with the semi-empirical color-effective temperature and bolometric corrections used in 
\cite{bell:2013}.
}
\label{fig:cmd_age}
\end{figure}

\begin{figure}
\includegraphics[width=3.2in]{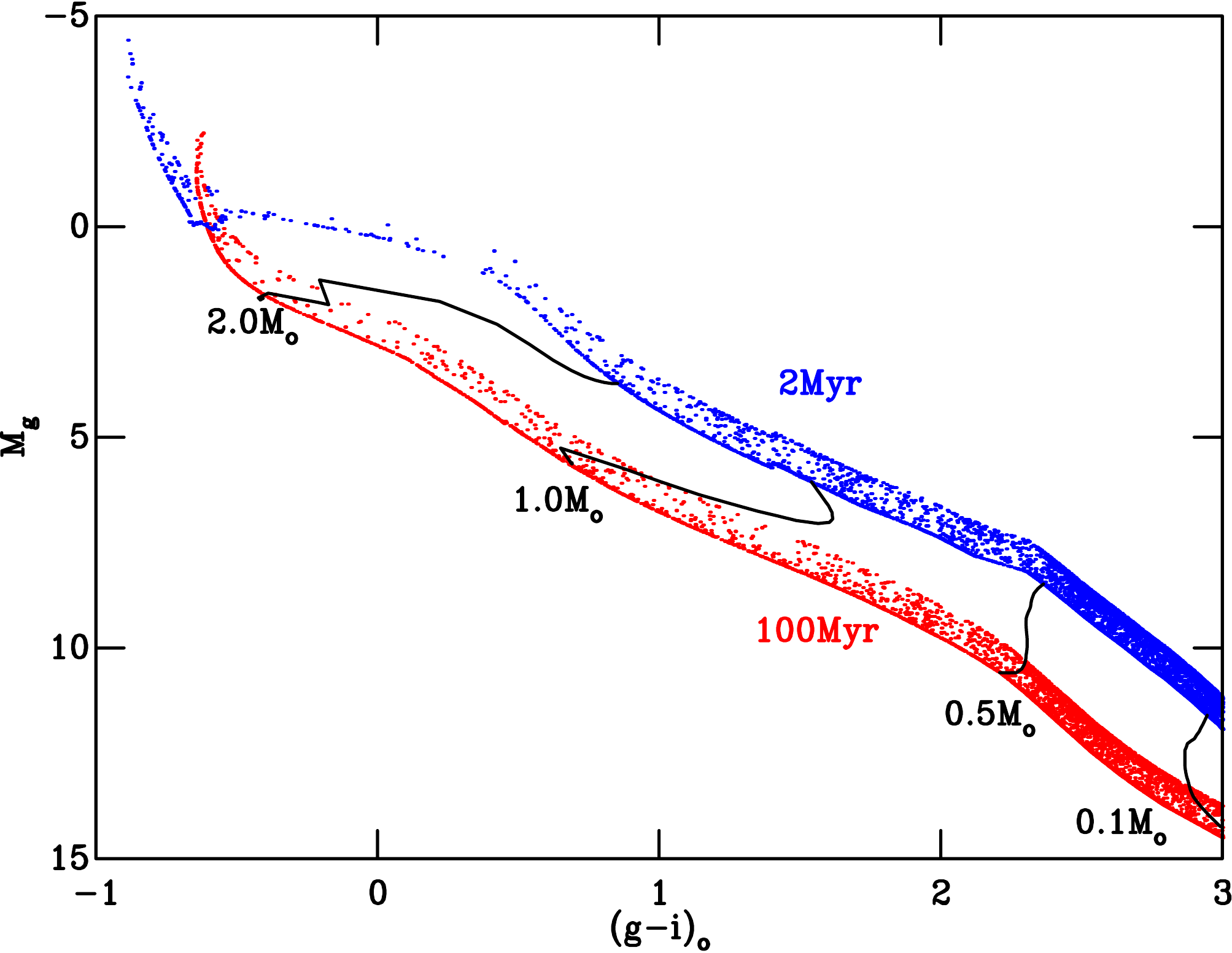}
\caption{A simulated color-magnitude diagram for clusters 2 and 100 Myr old.
The masses are sampled from a Kroupa mass function \citep{dabringhausen:2008}, with the prescription for binaries as in \cite{bell:2013}.
The interior models used are (from high to low mass) \cite{schaller:1992},
\cite{dotter:2008} and \cite{baraffe:1998} and the model atmospheres from \cite{Castelli:2004} (hot stars) and \cite{allard:2011} (cool stars).
We have chosen to join points between the models to minimize the dislocations, though they are still (just) visible.
}
\label{fig:2age_cmd}
\end{figure}

\subsubsubsection{Transformations and Empirical Errors}

If the theory is transformed into the observational plane, it must be done carefully and systematically to ensure that the theory and the data are in the same photometric system.  Alternately, the observations can be transformed into the theoretical plane, requiring similarly detailed attention. Empirical requirements are a stellar spectral type or a direct estimate of the stellar temperature using model atmospheres, and at least two bands of photometry that enable a reddening estimate via comparison to the unreddened color expected for a star of the same spectral type or temperature.  While traditional spectral typing was done in the ``classical MK" region from 4000-5000 \AA, the significant extinction towards star forming regions led to the subsequent development of spectral sequences in the red optical and then the near-infrared \citep[summarized in][]{gray:2009}.

In practice, locating young PMS stars in the HRD involves many challenges.  One is that the effects of changing surface gravity as stars contract to the main sequence often are not considered with the rigor they deserve.  The mass-dependent surface gravity evolution over the tens to hundreds of Myr time-scales that it takes high-mass to low-mass stars to reach the main sequence affects temperatures, colors, and bolometric corrections.

An intermediate temperature scale has been advocated by \cite{luhman:1999} for stars a few Myr old having spectral types later than about M4, motivated in part by a desire to match the isochrones of \cite{baraffe:1998} that could be brought about by assuming the stars to be warmer.  Intermediate color scales in various bands have been investigated by \cite{lawson:2009,dario:2010,scandariato:2012,pecaut:2013}.   The differences from the main sequence are not necessarily systematic.  For example, some colors are redder towards lower surface gravity down to some spectral type, such as M3 for $(V-I)$, and then become bluer than main sequence values continuing to later types.  There is also need for consideration of intermediate bolometric corrections rather than the broad application of main sequence relations, which are increasingly incorrect towards later spectral type young PMS stars. Figure~\ref{fig:cmd_age} shows an example of intrinsic color and bolometric correction differences on transformation of the same evolutionary tracks. 

Another challenge in placing young stars in the HRD is the extinction correction.  Reddening towards young populations is often differential, that is, spatially variable across a star forming region with some or most of the reddening effect possibly arising in the local circumstellar environment itself.   Thus the extinction corrections must be performed on a star-by-star basis.  Furthermore, a wavelength range must be found for determination of the reddening correction and application of the bolometric correction that is dominated by stellar (as opposed to strongly contaminated by circumstellar) emission.

Many young stars exhibit excess flux due to either or both of: accretion luminosity, which peaks in the ultraviolet but can extend through the entire optical wavelength range out to $Y$-band,  and disk emission, which peaks in the mid- to far-infrared but can extend as short as $I$-band.  The appropriate wavelength range for de-reddening appears to be in the blue optical ($B$, $V$) for earlier type stars, in the red optical ($I$, $Y$) for late type stars, and possibly in the near-infrared ($J$) for young brown dwarfs.  An alternate approach is to use high dispersion spectroscopy to determine a specific veiling value (the wavelength-dependent excess continuum flux due to accretion relative to the stellar photospheric flux) in a particular band, and use it to veiling-correct the broad band photometry before dereddening or application of a bolometric correction.

The presence of disks and accretion adds a further complication in the form of photometric variability.   This occurs at the level of a few percent for more active young spotted stars, up to the 5-30\% levels typical of stars with disks that are accreting.  Unless the true nature of the variability is understood, i.e., whether it is caused mostly by accretion effects or mostly by high latitude dust obscuration events, it is not clear whether the right choice for placing stars in the HRD would utilize the bright-state vs the faint-state magnitudes from a data stream.  For disks with either large inclination or scale height, the observed photometry may be dominated by scattered light.  This can render the stars fainter than they would be if seen more directly, and also leads to underestimates of the extinction.

The minimum statistical errors on just the basic parameters required from observations for HRD placement are $\sim$50-200K in $T_{\rm eff}$ and $\sim$0.01-0.1 mag on measured photometry, along with $\sim$10\% uncertainty in mean cluster distance.   An additional systematic error is the usually unknown multiplicity status of individual sources observed in seeing-limited conditions; this has an effect that depends on the binary mass ratio  with maximum amplitude of 100\% in luminosity overestimate for an equal mass/age system (see, e.g., the simulations at two ages Figure~\ref{fig:2age_cmd}), modulo any error due to differential extinction.

Likely additional random errors for the youngest stars in star-forming regions where disk effects may complicate derivations, include a typical $\sim$0.05 mag in average photometric variability  and $\sim$0.3-1 mag in visual extinction determination if optical colors are used, or 2-5 mag if infrared colors are used (not including a possible systematic effect due to the form of the reddening law in environments known to be exhibiting grain growth). In practice, the effective extinction errors are somewhat lower since HR diagrams are generally made by applying bolometric corrections to $I$-band or $J$-band photometry. For pre-main sequence stars there is also a 0.05-0.1 mag error from intrinsic color uncertainty, another 0.05-0.1 mag from the bolometric correction uncertainty, plus an additional $<$10\% uncertainty on individual stellar distances due to cluster depth (at 150pc, assuming the clusters are as deep as they are wide on the sky, though less for more distant star forming regions).  

The above numbers suggest $\sim$ 0.2-0.3 dex in random error alone, driven by the error in the extinction correction, to which the systematic distance and binary errors would need to be added.  If this is the case, summation of all error sources is roughly consistent with, though arguably somewhat less than \citep{burningham:2005}, the 0.2-0.6 dex rms spread in luminosity that can be measured for most young clusters, a topic we discuss in detail in Section \ref{sec:agespread}. From similar considerations, \cite{hartmann:2001,reggiani:2011} and \cite{preibisch:2012} each estimate $\sim$0.1-0.15 dex as the typical luminosity error, in which case the observed luminosity spreads are much greater than that attributed to random error. However, as illustrated recently by \citet{manara:2013}, there are large trade-offs between accounting for accretion and extinction in analyzing observed spectra and colors.  Thus the propagation of random errors may grossly underestimate the true luminosity uncertainties of heavily extincted accreting stars, which can be subject to large systematics.  These issues possibly account for the factors of several variation among authors in the extinctions and luminosities reported for the same young stars. For populations that are more distant than the closest regions and/or for which differential extinction and the uncertainties induced by accretion/disk effects are smaller, the errors in HRD placement would be lower, $<$0.1-0.15 dex uncertainty in luminosity to accompany the 50-200K uncertainty in temperature. 

Simulations in, e.g., \citet{hillenbrand:2008,naylor:2009,preibisch:2012,bell:2013} quantify how, when the multiplicity effect dominates, stars appear systematically more luminous and hence younger in the low mass pre-main sequence HR diagram.  To quantify, considering most of the applicable error terms, \citet{preibisch:2012} reports simulating that a true 2 Myr old star had inferred ages ranging over 1.2-2.2 Myr (1$\sigma$) around a mean age of 1.7 Myr, while a true 5 Myr old star had inferred ages ranging over 2.7-5.5 Myr (1$\sigma$) around a mean age of 4.1 Myr.  Derived stellar ages in studies that do not account for multiplicity and known sources of error would appear young by 15-20\%, but also would scatter by factors of nearly 50\%.  At older ages, the dominating systematic binary effect corresponds to a much larger age difference than at the younger ages, given the spacing of the isochrones. In the youngest regions where the extinction and disk effects discussed above dominate over the multiplicity effect, the luminosity errors may be more symmetric and hence the ages not necessarily biased low.

\subsubsubsection{Star Forming Region and Cluster Results}

\begin{figure}
\includegraphics[angle=-90,width=3.5in]{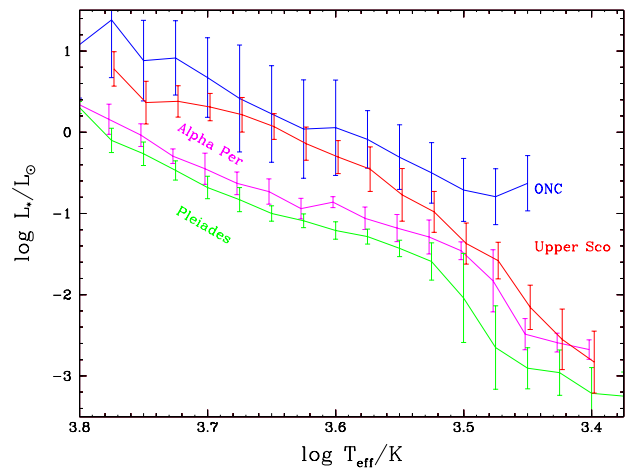}
\caption{Median luminosity and luminosity dispersion exhibited by four example benchmark clusters:
the Orion Nebula Cluster, Upper Sco, $\alpha$ Per, and the Pleiades.  Error bars indicate
the empirical 1$\sigma$ spread in luminosity with the red and magenta bars artificially offset 
along the abscissa for clarity.  Stars were placed in the HR diagram 
using available spectral types and optical photometry from the literature, and the 
median luminosities calculated at each 0.025 dex bin in temperature.
}
\label{fig:hrd_clusters}
\end{figure}

Figure \ref{fig:hrd_clusters} illustrates empirically how inferred stellar luminosities evolve 
from very young star forming regions associated with dense gas, like the Orion Nebula Cluster, 
to older dispersing young associations lacking molecular gas, like Upper Sco, 
to open clusters like $\alpha$~Per and the Pleiades.  As expected based on the error
considerations discussed above, the luminosity spreads are larger and more symmetric
at younger ages but settle down to the predicted $\pm$0.1-0.2 dex of scatter at older ages,
with the asymmetric signature of the binary effects apparent in most clusters older than $\sim$15 Myr.
Median luminosity plots like these can provide an age ordering to clusters and star forming
regions -- as long as the distances are accurately known and the samples unbiased.  
For example, it is possible to rank in increasing age (though not necessarily distinguish 
the ages of individual members, due to the apparent luminosity spreads)
e.g. NGC~2024 and Mon~R2, followed by the ONC, then Taurus and NGC 2264, IC~348 and $\sigma$~Ori,
Lupus and Chamaeleon, TW Hydra and Upper Sco, UCL/LCC, IC~2602 and IC~2391, $\alpha$~Per, 
and finally the Pleiades cluster.  
Due to the general luminosity decline with age for low mass pre-main sequence stars, 
this ranking seems meaningful even if the absolute ages remain highly debated, 
and even though the run of luminosity with effective temperature for any given cluster 
is not well-matched by modern theoretical isochrones.  

Performing the same exercise in color-magnitude diagrams, instead of the HR diagram, \citet{mayne:2007} and \citet{bell:2013} concluded that empirical tuning of bolometric corrections and temperature scales to particular models is necessary \citep[see also][]{stauffer:2003} in order to derive absolute ages.  Age ranking avoids this problem, and that of luminosity spreads due to multiplicity.  Combining \citet{mayne:2007} and \citet{bell:2013} we find that the ONC, NGC~6611, NGC~6530, IC~5146 and NGC~2244 are indistinguishable in age, but are the youngest clusters.  Similarly there is a somewhat older group composed of $\sigma$~Ori NGC~2264, IC~348 and Cep OB3b.  Beyond this age a ranking is possible for individual clusters, with $\lambda$~Ori next, followed by NGC~2169, NGC~2362, NGC~7160, $\chi$~Per, and finally NGC~1960
at $\sim$20 Myr. 

In addition to testing whether empirical cluster sequences match theoretical models
tied to stellar age, comparisons can be made of the coevality in HRDs of 
spatially resolved pre-main sequence binaries \citep[e.g.][]{hartigankenyon:2003, kraushill:2009} as well as of the radii 
inferred from double-line eclipsing pre-main sequence binaries.  Such eclipsing systems
along with astrometric systems also enable fundamental mass determinations
for comparison with theory \citep[e.g.,][]{gennaro:2012}. Given that there are still 
systematic discrepancies between models and fundamental masses in the pre-main sequence phase
\citep[e.g.][]{hillenbrand:2004}, the ages predicted by these same models likely
also suffer systematic errors.

\smallskip
\noindent{\bf Summary of isochrone placement for PMS stars:}
\begin{description}
\setlength{\itemsep}{-4pt}
\item[$+$] Stellar luminosity changes rapidly and systematically during pre-main sequence contraction.
\item[$-$] Many factors influence the apparent luminosity of a PMS star, with not all of them known to us.  In addition, the resulting change in luminosity with time may not be monotonic.
\item[$-$] The differences in published models are large.
\item[$-$] As a result of the above points, absolute ages from PMS isochrones are not yet credible. 
\item[$-$] Placing individual stars in an HRD requires good estimates of $T_{\rm eff}$ and $L_{\rm bol}$, both of which can be affected by extinction and non-uniform reddening, and also circumstellar material and accretion.
\item[$+$] When the above effects are small, CMDs may be sufficient, and indeed more precise given the quantification of $T_{\rm eff}$ inherent in spectral typing.
\item[$-$] Unresolved binaries add uncertainty to luminosities.
\item[$-$] Estimating errors is difficult because of poorly-understood factors that lead from the observations to modeled quantities.
\item[$-$] The effects of observational uncertainties increase the age errors towards the ZAMS as the evolution slows and the isochrone density increases.
\end{description}


\subsection{Main-sequence evolution to and beyond the turn-off}  

Although the majority of stars in young clusters and star-forming regions are still in their PMS phase, a small number of the more massive stars will have reached the main-sequence, and indeed the most massive will have evolved beyond beyond it.  The position of these stars in either the CMD or HRD gives us more age diagnostics which are based on very different physics from PMS evolution.  Terminology is important here.  While the star is still undergoing core hydrogen burning it is (by definition) evolving from the ZAMS to the terminal-age main sequence (TAMS), and an age derived from this gradual cooling and increase in luminosity is best described as a {\it main-sequence age}, or sometimes a {\it nuclear age}.  Once a star reaches the turn-off, the evolution is rapid, and so a further possible age diagnostic is the luminosity of the top of the main-sequence; a {\it turn-off age}.  Finally, the position of a star in its post-main-sequence evolution can be yield useful ages, though the paucity of these high-mass stars can cause difficulties.

Overlaying main-sequence isochrones on the upper-main-sequence can yield useful constraints on the age of a cluster \citep{sung:1999}, allowing one to match both the
position of the turn-off and the main-sequence evolution.
However, more precise ages with statistically robust uncertainties, can be obtained by using modern CMD fitting techniques \citep[e.g.,][]{naylor:2009}.

A clear advantage of this part of the CMD/HRD is that the model atmospheres are well understood, lacking the complications of molecular opacities.  The potential problems lie in the interior models, and involve enhanced convective overshoot (whether by rotation or otherwise) and mass loss.  Stellar rotation will induce better mixing, enabling a star to burn hydrogen longer, but it also makes a star less luminous in its early MS evolution \citep{ekstrom:2012}, and so the overall position of the isochrones changes little.  As illustrated in \cite{pecaut:2012}, the calculated age for Upper Sco changes by only 10 percent between rotating and non-rotating isochrones.   Similar arguments apply to enhanced convective overshoot, with a similarly small effect \citep[compare Figs.~4 and 5 of][]{Maeder:1981}. Mass loss also tends to move the masses along the isochrone, and again has only a small effect on the overall shape and position  \citep[see the discussion in][]{naylor:2009}.  However, the position of the turn-off does change significantly with rotation, a given luminosity corresponding to an age of $\sim$30 percent younger at 10 Myr for the rotating case. 

In summary, main-sequence fitting seems to offer a method where the model dependence is small, but the model dependence for the turn-off is stronger.  However, the movement from the ZAMS to the TAMS is subtle, and requires either well calibrated photometry, or excellent conversion into the HRD.
Furthermore, right at the top of the main-sequence, which is the part most age sensitive part, is exactly where the mass function can push us into small number statistics, which also makes the technique vulnerable to uncertain extinction and binarity.

\cite{lyra:2006} showed that for the range 10-150~Myr, main-sequence ages and PMS ages agree.  However, \cite{naylor:2009} showed that for clusters younger than 10~Myr the main-sequence ages were a factor of two larger than those derived from PMS isochrones.  \cite{pecaut:2012} showed that there is excellent agreement between the isochronal ages ($\sim$11 Myr) for Upper Sco members that were post-MS (Antares), MS B-type stars, and PMS AFG-type stars; lower mass stars, however trend towards younger ages. \cite{bell:2013} show reconciliation of the MS and PMS ages for several well-studied clusters, with a general movement towards higher ages than most previous studies. Reasonable concordance between MS, PMS and LDB ages has been demonstrated for the oldest ($22\pm 4$\,Myr) cluster in the Bell et al. sample \citep{jeffries:2013}. 

\smallskip
\noindent{\bf Summary of isochrone placement for MS stars:}
\begin{description}
\setlength{\itemsep}{-4pt}
\item[$+$] Post-ZAMS models are in better agreement with one another and with the data, compared to PMS models. 
\item[$-$] Main sequence evolution is slow and higher mass stars rare, leading to errors in inferred age that are driven mostly by the observational uncertainties and small number statistics.
\end{description}


\subsection{Surface gravity diagnostics}  
\label{sec:gravity}

The gravities of stars of the same temperature, but spanning ages from 1 to 100 Myr can be different by 1 dex in surface gravity, depending on the temperature range considered \citep[e.g.,][]{dotter:2008}.  This will lead to differences in both the spectra and the colors that in principle could be used to determine the gravity and hence, in combination with isochrones, the age.  However, the differences in color due to gravity for a given $T_{\rm eff}$ are modest for temperatures above 3000 K. 
Furthermore, surface gravity effects would have to be disentangled from reddening effects when interpreting observed colors. 

Spectroscopic measures of surface gravity are more promising and have the key advantage of distance independence. Specifically, there is a ``triangular" shape of the $H$-band spectra of low-gravity late-M stars \citep{lucas:2001, allers:2007} which, although used as a method of selecting young stars, is not a viable route to precise ages.  Other gravity-sensitive lines include the K-band Na I 2.206 $\mu$m line \citep{takagi:2011} and the J-band K I doublets at 1.17 and 1.25 $\mu$m \citep[e.g.][]{slesnick:2004}.

The optical Na 8190 \AA\ doublet is now widely used for mid- and late-M stars.  Line strength is driven by both gravity and temperature, so two-dimensional classification is required \cite[e.g.,][]{slesnick:2006, lawson:2009}.  \cite{schlieder:2012} compare data with theoretical predictions and conclude that although this line, like the infrared lines, clearly can be an effective diagnostic for identifying young stars, it may not be particularly useful for accurate age-dating of individual stars. However, it appears to do a good job of sorting cluster samples \citep{lawson:2009} with a resolution of a few Myr, and fitting of model atmospheres by \citet{mentuch:2008} leads to an association age ranking from youngest to oldest as: $\eta$ Cha, TW Hya, $\beta$ Pic, Tuc/Hor, and AB Dor.
\citet{prisinzano:2012} have examined the log g sensitivity of the Ca I lines around 6100 \AA, which appear useful for late G, K, and M type stars.  Bluer surface gravity sensitive lines such as the Mg I b triplet are generally not practical for faint, extincted young pre-main sequence stars.

\subsection{     
Pulsations and seismology
}
\label{sec:seismology}

Solar-like p-mode oscillations have now been detected in hundreds of stars with the ultra-precise photometry of the CoRoT and Kepler missions (e.g., \citet{chaplin:2013}).  These include many evolved stars, in which it is now possible to distinguish stars ascending the red giant branch from those descending, even though they occupy identical regions in the HRD (\citet{bedding:2011}).  Observations of stellar oscillations can reveal interior properties that constrain model fits much more precisely than is possible from only surface information. 

The p-mode oscillations have also been detected in stars near one solar mass, but the detections are mostly for older stars more massive than the Sun \citep{chaplin:2013}.  When detected, these oscillation frequencies provide significant physical constraints (e.g., density vs. radius)on stellar models, enabling ages to be calculated to as precise as 5-10\% if spectra have been obtained to establish the star's composition.

Similar oscillations have not been detected in PMS stars, although there have been few attempts.  The primary reason is the inherent high level of photometric variability in very young stars that is related to their high activity levels \citep{chaplin:2011}.  This leads to a noise level that inhibits detecting the oscillations even if they are present.  However, some of the more massive PMS stars cross the instability strip in the HRD as they approach the main sequence.  These intermediate-mass stars include  $\delta$ Scutis and some B stars that oscillate via the $\kappa$ mechanism \citep*[e.g.,][]{ruoppo:2007, alecian:2007}.  Pulsations of this kind have been seen in some stars of NGC 2264 \citep{zwintz:2013}, and may be capable of providing model-dependent ages that are more precise than those from the HRD as they are distance-independent and insensitive to extinction.

Pulsations are also predicted for low-mass stars and brown dwarfs via the $\epsilon$ mechanism \citep{palla:2005}, though thus far the amplitudes are not well constrained theoretically and only upper limits are available from observations \citep{cody:2011}.

\smallskip
\noindent{\bf Summary of seismology for PMS stars:}

Lower-mass stars, which form the vast majority of PMS stars, appear to be too noisy to exhibit detectable oscillations.  Some higher-mass stars fall in the instability strip and so detailed observations of them can provide physical constraints that help lead to more precise ages.


\subsection{     
Projected stellar radii
}
\label{sec:rsini}

Young stars with convective envelopes are magnetically active and often rapidly rotating (section~\ref{sec:rotation}; see also the chapter by Bouvier et al.). These properties combine to produce light curves that are rotationally modulated by cool starspots, from which rotational periods, $P_{\rm rot}$, can be estimated. The projected equatorial velocity of a star, $v \sin i$ where $i$ is the inclination of the spin axis to the line of sight, can be estimated using high resolution spectroscopy to measure rotational broadening. If these quantities are multiplied together, the projected radius of a star is ($R/R_\odot) \sin i = 0.02 (P_{\rm rot}/$day$) (v\sin i/{\rm km\,s}^{-1})$. 
Although $i$ is unknown, this formula gives the minimum radius of the star and because PMS stars become smaller as they move towards the ZAMS, then this minimum radius can be compared with isochrones of $R$ versus $T_{\rm eff}$ to give a model-dependent upper limit to the age.

This method is independent of distance and insensitive to extinction and binarity (as long as one star is much brighter), but requires time series photometry, a spectrum with sufficient resolution to resolve the rotational broadening and an estimate of $T_{\rm eff}$. For field stars this can be a useful technique to confirm their youth, with a range of applicability equal to the time taken to reach the ZAMS at any particular spectral type \citep[e.g.,][]{messina:2010}. In large groups of stars where spin axis orientation can be assumed random, then the same technique offers a way of statistically determining the average radius or distribution of radii (see section~\ref{sec:agespread}). The main disadvantage of the method is that while it offers a geometric radius determination, independent of any luminosity estimate, interpreting this in terms of an age is still entirely model dependent. Furthermore, there are strong indications that the radii of young, magnetically active stars are significantly larger than predicted by current models, and this could lead to a systematic error \citep{lopez:2007,jackson:2009}.


\subsection{     
Location and size of the radiative-convective gap
}
\label{sec:rcgap}

It has long been known that there is a discontinuity in the CMDs of young clusters at an age dependent mass. This is illustrated in the 2\,Myr model of Figure \ref{fig:2age_cmd} where there is a gap at $(g-i)=0$ with MS stars to the blue, and the PMS stars to the red \citep[see][for an observational version]{walker:1956}.  The gap is created because there is a rapid evolution between the two sequences, as the stars move from the (largely) convective pre-MS, developing radiative cores (a progression along the Henyey tracks). Various authors have recognized this resulting gap, calling it the ``H feature'' \citep{piskunov:1996}, the ``Pre-Main-Sequence Transition" \citep{stolte:2004} and the ``radiative-convective" or R-C gap \citep{mayne:2007}.  Both \cite{gregory:2012} and \cite{mayne:2010} discuss the gap in its wider context, but for age determination its importance stems from the fact that its position and size vary as a function of age.  It becomes almost imperceptible after 13\,Myr \citep[see the figures of][]{mayne:2007}.

The position of the gap in absolute magnitude can be used as an age indicator since it produces a dip in the luminosity function.  This is particularly useful in the extra-galactic context where distances may be known from other information, where this method has been pioneered by \cite{cignoni:2010}, but has also been applied to Galactic clusters \citep{piskunov:2004}.

The real potential power of the R-C gap, however, is that at a given age its width measured in either color or magnitude space is independent of both the distance and the extinction.  The practical problem is observationally defining the edges of the gap, there being issues of both field star contamination, and the fact that there clearly are a few stars in the gap itself \citep{rochau:2010}.  Given a clean sample, perhaps from Gaia, one may be able to solve this problem using two dimensional fitting to model isochrones \citep[see][]{naylorjeffries:2006}, but whether the models can correctly follow the rapid changes in stellar structure through this phase remains to be tested.  An alternative may be to construct empirical isochrones.


\section{
EMPIRICAL METHODS}

All the empirical methods for determining stellar ages are inherently circular in their reasoning, especially for PMS stars.  Indeed, the general trends and variations in properties is often the very focus of investigations.

\subsection{
Rotation and Activity in PMS stars      
}
\label{sec:rotation}

Using the observed rotation rates of PMS stars to estimate their ages conflicts with one of our primary goals, which is to establish ages independently of the phenomena being studied.  The origin and evolution of angular momentum -- and its surface observable: the rotation period or $v \sin i$ -- is one of the major areas of study in the field of star formation and early stellar evolution (see Bouvier et al. in this volume).

Nevertheless, we ask here if our knowledge of rotation and how it changes is sufficient to invert the problem to estimate an age.  And if the process will not work for a single star, is there some sample size for which a median or average rotation is a clear function of age?

The essence of the situation is illustrated in Figure \ref{GB12}, which is taken from \citet{gallet:2013}.  The colored lines show various models which are not of interest here.  Instead, note the range of observed rotation periods for solar-mass stars in a number of young clusters.  First, in any one cluster the range is 1.5 to 2.5 dex.  Second, this spread is not ``noise'' to which Gaussian statistics apply because the values are not concentrated toward an average value but instead are spread fairly uniformly in $\log P_{\rm rot}$.  Third, the data for any one cluster encompasses a range of masses, but for these young stars $P_{\rm rot}$ varies little with mass (or color), so that does not matter.  Fourth, and most important here, for the first $\sim100$ Myr or so there is no clear trend at all in mean or median rotation.  Even taking 10 or 20 stars to average would not lead to a useful result.  In terms of rotation, a group of stars at 10 Myr looks very much like a similar group at 100 Myr.

This lack of distinct change in the distributions of $P_{\rm rot}$ values is in part due to the systematic decrease in stellar radii over the same age range as these stars approach the ZAMS: angular momentum loss is balanced by the star's shrinkage to keep $P_{\rm rot}$ roughly constant. At lower masses it seems that angular momentum loss may not be so efficient. \citet{henderson:2012} point out that between 1 and 10\,Myr the fraction of slowly rotating M-stars in young clusters rapidly diminishes. This empirical finding will not yield reliable ages for individual M-dwarfs, because a slowly rotating object could be either very young or quite old. In groups of (presumed coeval) M-dwarfs, then unless distances are unavailable then PMS isochrones in the HRD/CMD are likely to lead to more precise ages.

\begin{figure}
\includegraphics[width=3.2in]{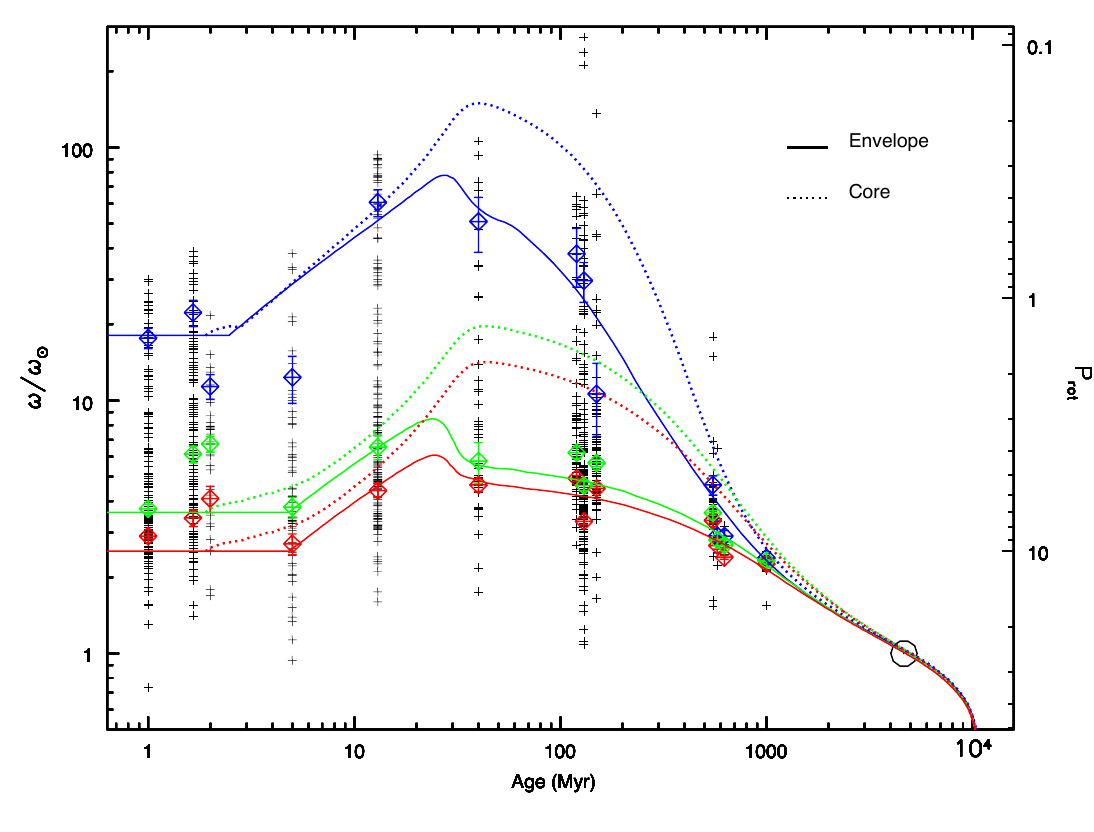}
\caption{Observed rotation rates of stars near one solar mass in young clusters \citep{gallet:2013}.  (See also the chapter in this volume by Bouvier et al.)  The colored lines show several models that are not of relevance here.  The plus symbols show observed rotation periods.  The blue, red, and green diamonds represent the 90th percentile, 25th percentile, and the median for each cluster's distribution of $P_{\rm rot}$.  Note the very large spread (1.5 to 2.5 dex at any one age) and the lack of a clear trend for at least the first $\sim 100$ Myr.}
\label{GB12}
\end{figure}

``Activity'' refers to various non-thermal emissions that arise in the upper atmospheres of stars.  The easiest to measure are the emission reversals in the cores of the Ca {\sc ii} H and K lines, but in T Tauri stars especially H$\alpha$ emission is prominent and a defining characteristic of the class.  Additional features are seen in ultraviolet wavelengths, and coronal emission in x-rays is often used as a means of identifying very young stars in a field because the x-rays are able to penetrate the surrounding material better than other wavelengths.

Activity in PMS stars is highly variable (as it is even for our Sun), at a level that masks any underlying age trend.  In addition, the activity seen in young stars tends to saturate \citep[e.g.,][]{berger:2010} so that there is little variation in activity with age up to $\sim$200 Myr or so, even in the mean for groups of stars.  There is even some evidence for ``super-saturation'' in which extremely active stars exhibit less emission in activity signatures such as x-rays compared somewhat less active stars \citep{berger:2010}.

Using activity to estimate age even for main sequence stars can be problematic.  There is a well-defined relation in the mean \citep[e.g.,][]{soderblom:1991} but substantial scatter within coeval groups and disagreement between members of binaries \citep{mamajek:2008b}.

\smallskip
\noindent{\bf Summary of rotation and activity for PMS stars:}

The inherent dynamic range of rotation and activity at any one age for PMS stars equals or exceeds any broad trend in average rotation rates and activity levels with time. In addition, activity saturates at the high rotation rates seen in PMS stars and the early rotation history appears to be tied to the presence of disks.  Both rotation and activity may be a useful age indicators for older pre-main and main sequence stars, depending on the mass (e.g., $\sim$0.5 Gyr and older for solar mass stars). Ages from seismology provided by {\it Kepler} are helping to calibrate these relations.  


\subsection{    
Accretion and circumstellar disks
}

\label{sec:disks}

At the earliest visible ages, most stars have near- to mid-IR excesses diagnostic of warm dust at 0.1-10\,au, which is thought to arise from a primordial circumstellar disk, and show signs of gas accretion. Observations in young clusters and associations then show how spectral energy distributions (SEDs) evolve and the fraction of stars exhibiting these signatures declines. Various age-dependent relationships have been proposed. The fraction of stars with excesses at near-IR wavelengths halves in about 3\,Myr and becomes close-to-zero after 10\,Myr \citep{haisch:2001, hillenbrand:2005, dahm:2007, hernandez:2008, mamajek:2009}. The median SED of class II PMS stars appears to show a monotonic progression with age \citep[e.g.,][]{alves:2012}. Likewise, the fraction of PMS stars showing accretion-related H$\alpha$ emission declines from around 60\% at the youngest ages, with an exponential timescale of $\sim$2--3\,Myr and becomes very small at $\geq 10$\,Myr \citep{jayawardhana:2006, jeffries:2007b, mamajek:2009, fedele:2010}. In principle these relationships, calibrated with fiducial clusters, could be inverted to obtain rough age estimates for stellar groups, or to rank the ages of populations with measured disk diagnostics.

In practice, there are complications that must be considered.  First, the diagnostics used must be defined and measured in the same way in the calibrating clusters and using stars of similar mass. There is evidence that the fraction of stars with IR-excess is both wavelength dependent and mass dependent (longer wavelengths probe dust at a larger radii and provide greater contrast with cool photospheres, lower mass stars have lower mass disks) \citep{carpenter:2006, hernandez:2010}. There are also differing methods of diagnosing active accretion which could alter the accretion disk frequency \citep{barrado:2003, white:2003}. It is also appears that disk dispersal timescales increase with decreasing stellar mass \citep{kennedy:2009, mamajek:2009, luhman:2012}.  
Second, an individual star may have a disk lifetime from $<1$\,Myr to $\sim 15$\,Myr. It is also possible that different clusters at the same age have different disk frequencies \citep{mamajek:2009}.  The mechanisms by which disks disperse and accretion decays are poorly understood and must involve parameters other than age, such as initial disk mass, exposure to external UV radiation, planetesimal formation, etc. \citep{adams:2004, williams:2011}. Until their relative importance is established, age estimates for groups of stars, and especially age estimates for individual stars, based on disk properties are at risk of significant error. e.g. A group of stars emerging from a molecular cloud that all possess near-IR excesses may be very young or perhaps they have been shielded from the disruptive influence of external UV radiation? Third, some diversity of disk/accretion diagnostics may be due to disk geometry or accretion variability. This should not be an issue for large statistical samples but is of serious concern for individual stars. Finally, disk frequency can only be considered a secondary age indicator at best. It requires calibration using clusters with ages determined by other methods and shares any inaccuracy with those methods. As we have seen in previous sections. no absolute age scale has been established on the timescales relevant for disk dispersal.

\citet{mamajek:2009} quantified the time evolution of the protoplanetary disk fraction in a group using the simple relation $f = \exp(-\tau/\tau_{\rm disk}$), where $f = n/N$ = number of stars with accretion disks $n$ divided by total number of stars $N$. With the data then available, $\tau_{disk}$ was estimated to be $\sim$2.5 Myr, but this depends on the ages adopted for the calibrating clusters and recent work suggests it could be a factor of two larger \citep[e.g.,][]{pecaut:2012, bell:2013}. There also appears to be real cosmic scatter in disk fractions at given age, and as a function of stellar mass and multiplicity.  The approximate {\it disk fraction age} of a group is thus:

\begin{equation}
\tau = -\tau_{\rm disk} \ln f = \tau_{\rm disk} (\ln N - \ln n)
\end{equation}

If one can quantify the intrinsic scatter $\sigma_{\tau}$ in the exponential decay constant $\tau_{\rm disk}$ (i.e., ``cosmic" scatter in the disk fraction due to the many factors that affect disk depletion), then the uncertainty in the age of a sample determined using disk fraction would be $\sigma$ $\simeq$ $\sqrt{ ({\rm ln} f)^2 \sigma_{\tau}^2 + \tau^{2}/n}$ assuming Poisson noise from the counting of disked stars. As an example consider the $\eta$ Cha cluster, where \citet{siciliaaguilar:2009} find that $n$ = 4 of $N$ = 18 members are accretors ($f$ = 22\%). Adopting a decay timescale $\tau$ = 2.5 Myr, intrinsic scatter $\sigma_{\tau}$ = 1 Myr (to be measured), one would estimate an age for $\eta$ Cha cluster of 3.8\,$\pm$\,2.0 Myr; i.e., the disk fraction gives an age estimate with a precision of $\sim 50\%$, but this precision becomes poorer when $f$ or $n$ are small.   This age is consistent with published estimates \citep{fang:2013a}.

\smallskip
\noindent{\bf Summary of circumstellar disks as an age indicator:}

There is a clear trend in protoplanetary disk fraction versus time which could be used as a rough age indicator, but better precision is likely to be available from other methods (e.g., PMS isochrones).
Its utility may be confined to instances where representative censuses of accretors and class III objects complete to some mass can be made (e.g., joint infrared/X-ray surveys) and distances are uncertain. More work is needed to understand and quantify the factors responsible for the cosmic scatter in disk fraction as a function of age (and mass,  multiplicity,  environment, etc.).

\subsection{     
Lithium abundances
}
\label{sec:li}

\begin{figure}
\includegraphics[width=3.2in]{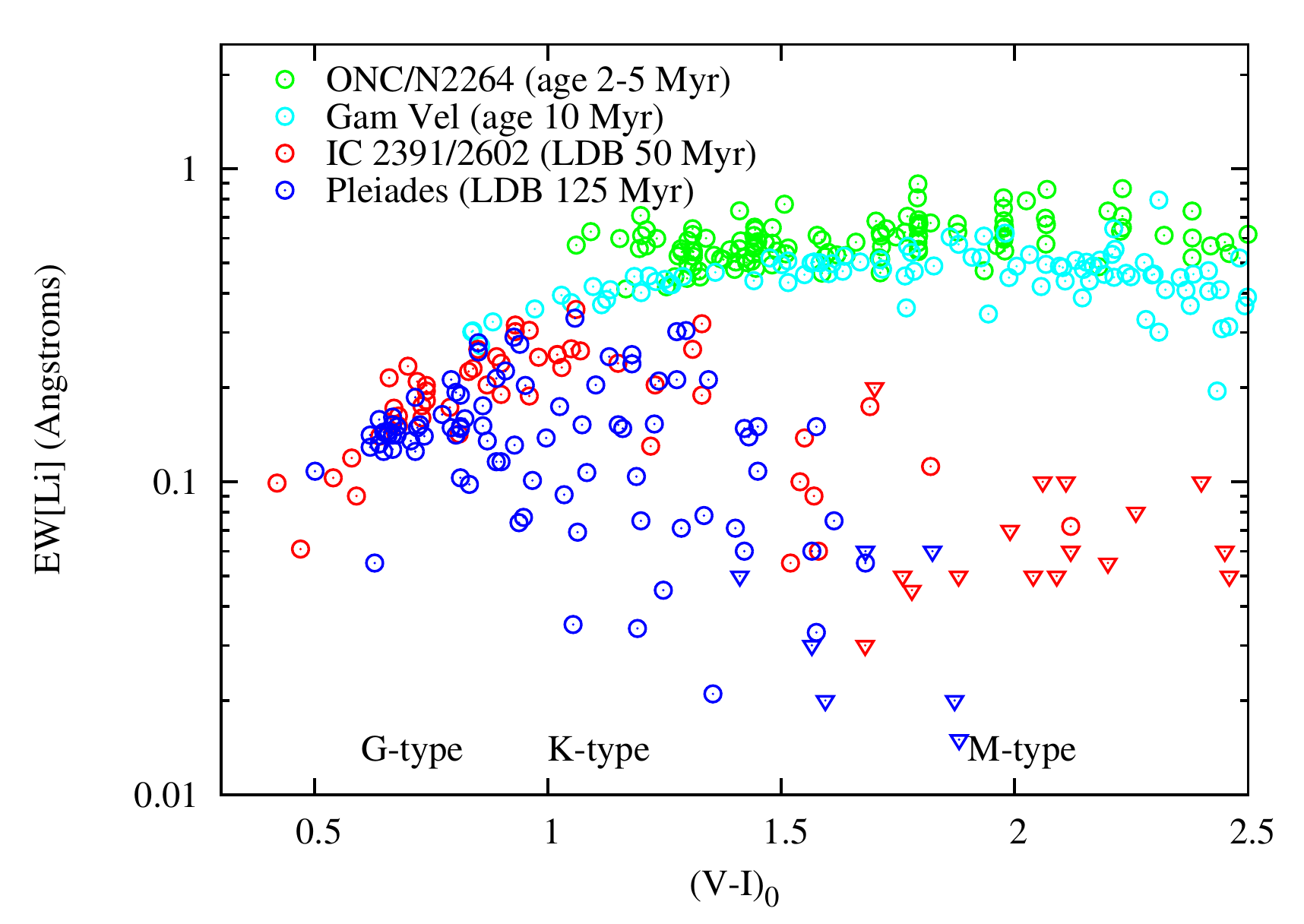}
\caption{The EW of the Li\,{\sc i}~6708\AA\ feature versus intrinsic color for groups of stars a fiducial ages. This demonstrates the empirical progression of Li depletion with age. Notable features are that: (i) The very young stars (2-10)\,Myr have EW(Li) consistent with zero depletion. (ii) Rapid depletion commences in M-dwarfs between 10 and 50\,Myr. (iii) Depletion occurs on longer timescales for K-dwarfs, but a large scatter develops. (iv) Little depletion occurs in the first 100\,Myr in G-dwarfs. Data are taken from \citet{soderblom:1993}, \citet{jones:1996}, \citet{randich:1997}, \citet{randich:2001}, \citet{sergison:2013} and the Gaia-ESO survey \citet{gilmore:2012}}
\label{fig:lidepletion}
\end{figure}

Li depletion in stars with $M>0.4\,M_{\odot}$
is more complex than in lower mass stars (see section~\ref{sec:ldb}). As the core temperature rises during PMS contraction, the opacity falls and a radiative core develops, pushing outward to include an increasing mass fraction. During a short temporal window, stars $\leq 1\,M_{\odot}$ can deplete Li in their cores and convectively mix Li-depleted material to the surface before the convection zone base temperature falls below the Li-burning threshold and photospheric Li-depletion halts. Li depletion commences after a few Myr in solar type stars and photospheric depletion should cease after $\sim 15$\,Myr, before the ZAMS, with a large fraction of the initial Li remaining. In contrast, the convection zone base temperature does not decrease quickly enough to prevent total depletion for stars of $0.4<M/M_{\odot}<0.6$ within 30\,Myr \citep[e.g.,][]{baraffe:1998, siess:2000}.  At masses $>1\,M_{\odot}$, the radiative core develops before Li-depletion starts and little PMS Li depletion can occur. These processes lead to predicted isochrones of photospheric Li depletion that are smoothly varying functions of mass or $T_{\rm eff}$ that could be used to estimate stellar ages.

Ages based on Li depletion among stars with $0.4<M/M_{\odot}<1.2$ (or equivalently late-F to early-M stars) are not ``semi-fundamental" in the same way as the LDB technique. Different evolutionary models make vastly different predictions; Li depletion is exquisitely sensitive to uncertain conditions at the base of the shrinking convection zone and hence highly dependent on assumed interior and atmospheric opacities, the metallicity and helium abundance and most importantly, the convective efficiency \citep{pinsonneault:1997, piau:2002}. It is also clear that ``standard" convective mixing during the PMS is not the only process responsible for Li depletion; the decrease in Li abundances observed between ZAMS stars in the Pleiades and similar stars in older clusters and the Sun is probably due to rotationally induced mixing processes or gravity waves \citep{chaboyer:1995, charbonnel:2005}.

Until stellar interior models are greatly improved Li depletion in low-mass stars can not provide accurate absolute ages. Indeed, Li depletion has chiefly been used to investigate the uncertain physics above using clusters of ``known" age. However, imperfect evolution models do not preclude using comparative Li depletion as an empirical age estimator or as a means of ranking stars in age order. Clusters of stars with ages determined by other techniques (LDB, MSTO) can calibrate empirical Li isochrones. There is a clear progression of Li depletion with age between the youngest objects in star forming regions, through young clusters like IC~2391/2602, $\alpha$ Per and the Pleiades (Fig.~\ref{fig:lidepletion}) to older clusters like the Hyades and Praesepe \citep*[sources of such data can be found in][]{sestito:2005}.  There are (at least) three physical sources of uncertainty in using these isochrones: (i) The initial Li abundance of a star is not known, but assumed close to meteoritic. The possible scatter may be estimated from Galactic chemical evolution models or from early G-stars in several young clusters at a similar age \citep[e.g.][]{bubar:2011}. It is unlikely to have a dispersion of more than around 0.1 dex among young population~I stars in the solar neighborhood \citep{lambert:2004}.  (ii) There is a scatter in Li abundance at a given $T_{\rm eff}$ in most young clusters. This is modest among G stars but grows to 1 dex or more in K-stars \citep[e.g.,][]{soderblom:1993, randich:2001}. The scatter may be due to mass-loss, rotation-induced mixing, inhibition of convection by magnetic fields or some other phenomenon \citep[e.g][]{ventura:1998, sackmann:2003}, but for determining the age of a particular star it is a major nuisance. (iii) Li depletion should be sensitive to metallicity and He abundance, particularly at lower masses. The range and availability of precise metallicities among available calibrating clusters is too limited to quantify this.

Effect (i) is only relevant for G-stars. The cosmic dispersion in initial Li may be significant compared to levels of Li depletion at $\leq 100$\,Myr. Effect (ii) will be more important in K and M-stars where Li is depleted by 1-2 dex in 100\,Myr. An individual star can be assigned an age but with an associated uncertainty that depends on the dispersion of Li abundance (or EW(Li), see below) in the coeval calibrating datasets. Effect (iii) means that at present, ages can only be properly estimated for stars with close-to-solar metallicity.

Efforts to estimate stellar ages using Li depletion have focused on its use as a relatively crude, but distance-independent, indicator of youth or as a means of supporting age determinations for isolated field stars or kinematic ``moving groups", where other age indicators are unavailable and HRDs cannot easily be constructed \citep{jeffries:1995, barrado:2006, mentuch:2008, brandt:2013}. The measurement of Li abundance in FGK stars requires high resolution ($R \geq 10\,000$) spectroscopy of the Li\,{\sc i}~6708\AA\ feature and the ability to estimate $T_{\rm eff}$ and $\log g$ from spectra or photometry. The abundances, derived via model atmospheres, are sensitive to $T_{\rm eff}$ uncertainties and also to uncertain NLTE effects (e.g. \citet{carlsson:1994}). If comparison with theoretical models is not required, then it makes little sense to compare data and empirical isochrones in the Li abundance, $T_{\rm eff}$ plane, where the coordinates have correlated uncertainties. A better comparison is made in the EW(Li), color plane (appropriately corrected for reddening), although there remain issues about the comparability of EW(Li) measured using different methods of integration, continuum estimation and dealing with rotational broadening and blending when the Li feature is weak or there is accretion veiling in very young stars.

Fig.~\ref{fig:lidepletion} illustrates how the effectiveness of Li depletion as an age indicator is dependent on spectral type. At very young ages ($<10$\,Myr) little or no Li depletion is expected and the observations are consistent with that. Between 10 and 50\,Myr, Li depletion is rapid in late K and M-dwarfs but barely gets started in hotter stars. Age values (as opposed to limits) in this range can only be estimated from EW(Li) for stars with these spectral types or for groups including such stars. M-dwarfs with Li (and below the LDB) must be younger than 50\,Myr. Conversely, M-dwarfs without discernible Li are older than 10\,Myr. Individual M-dwarfs cannot be age-dated much more precisely than about $\pm 20$\,Myr due to the wide dispersion of EW(Li) among the M-dwarfs of clusters at 40-50\,Myr \citep{randich:2001} and the lack of data for calibrating clusters at 10-40\,Myr. At older ages the focus moves to K-dwarfs with longer Li depletion timescales; these can be used to estimate ages in the range 50\,Myr (where some undepleted K-dwarfs are still found) to $\sim 500$\,Myr (where all K-dwarfs have completely depleted their Li). Within this range, the likely age uncertainties for one star are $\sim 0.5$\,dex, due to the spread in Li depletion observed in calibrating clusters, but there may also be a tail of extreme outliers and close binarity can also confuse the issue. The uniform analysis of clusters in \citet{sestito:2005} shows that the Li abundance drops by only $\sim 0.2$--0.3\,dex over the first 200\,Myr of the life of an early G star. With an observed dispersion 0.1--0.2\,dex, and similar uncertainties associated with metallicity and initial Li, then Li depletion cannot be used to estimate reliable ages for young G stars.

\smallskip
\noindent{\bf Summary for lithium depletion:}
\begin{description}
\setlength{\itemsep}{-4pt}
\item[$+$] The primary Li feature at 6708 \AA\ is easy to detect and measure in PMS stars, having an EW of 100 m\AA\ or more and lying near the peak of CCD sensitivity.
\item[$+$] The broad trend of declining Li abundance with age is strong for PMS stars, particularly late K- and M-dwarfs.
\item[$+$] PMS stars at $<10$\,Myr all appear to be Li-rich There may be exceptions \citep[e.g.,][]{white:2005}, but only in small numbers (see section~\ref{sec:agespread}).

\item[$-$] Detecting Li in a star requires spectra of good resolution and signal-to-noise.  Only one feature at 6708 \AA\ is generally visible.
\item[$-$] Li depletion is not understood physically; additional theoretical ingredients are likely to be required.  Models can reproduce the solar abundance \citep[e.g.,][]{charbonnel:2005} but there are no other old stars with well-determined Li abundances and fundamental parameters to constrain the theory. 

\item[$-$] Converting an observed Li EW to an abundance is very temperature sensitive, can require a non-LTE correction and may be influenced by surface inhomogeneities such as starspots \citep{soderblom:1993}. This and the item above prevent any absolute age determination from Li measurements.

\item[$-$] Substantial and poorly understood scatter exists in Li for stars of the same age, particularly below 1 $M_\odot$, which diminishes its effectiveness as an empirical age indicator.

\item[$-$] The rate of Li depletion is too low to be useful at ages $\leq 100$\,Myr and $M\geq 1\,M_\odot$.

\end{description}

\section{
AGE SPREADS AMONG PMS GROUPS
} 
\label{sec:agespread}

\begin{figure}
\includegraphics[width=3.2in]{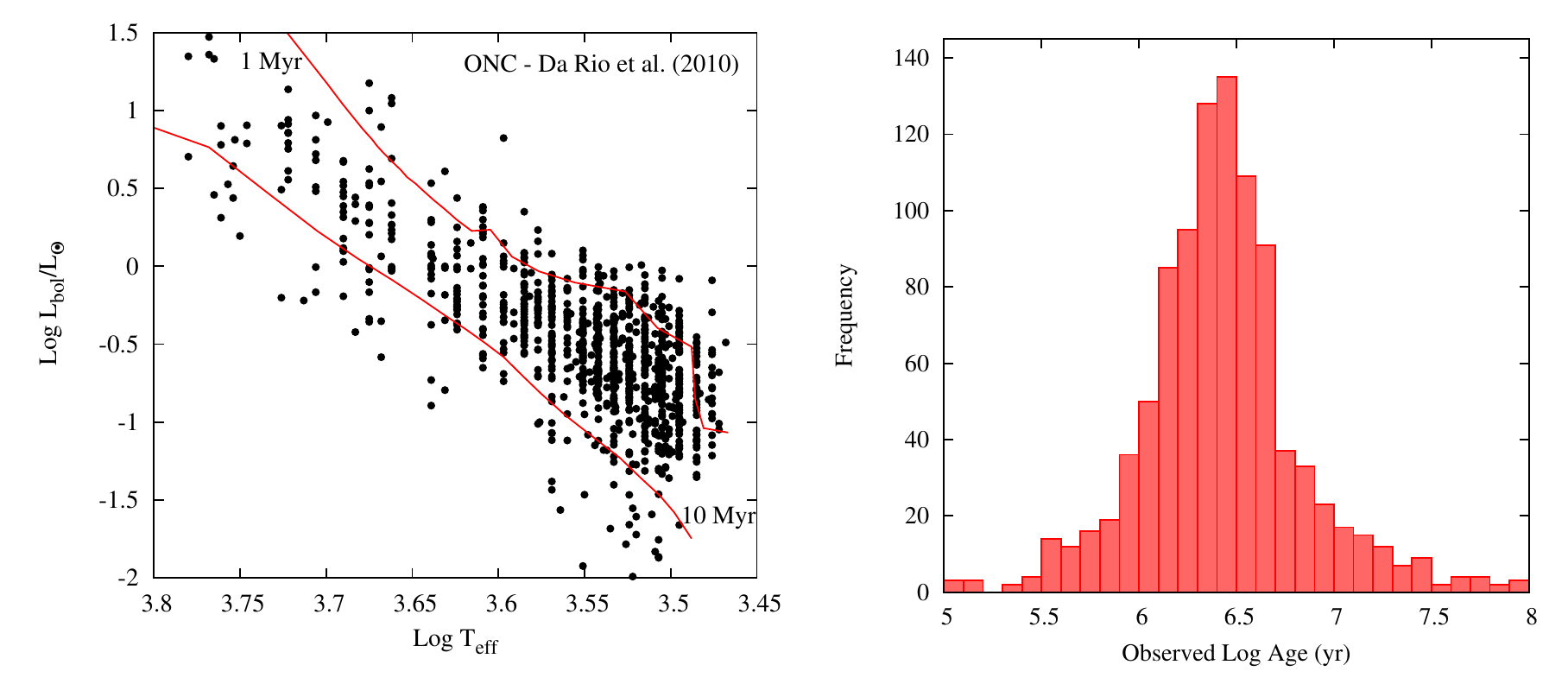}
\caption{(Left) The HR diagram of the Orion Nebula Cluster from the catalogue compiled by \citet{dario:2010}. The loci are isochrones at 1 and 10\,Myr from the models of \citet{siess:2000}. (Right) The inferred distribution of log age (in years) for these stars.
}
\label{fig:oncspread}
\end{figure}

There is considerable debate about how long it takes to form a cluster of stars. Some argue that the dissipation of supersonic turbulence in molecular clouds leads to rapid collapse and star formation on a freefall time ($<1$\,Myr for a cluster like the ONC: \citet{ballesteros:2007, elmegreen:2007}). Others take the view that turbulence is regenerated or the collapse moderated by magnetic fields and that star formation occurs over 5-10 freefall times \citep{tan:2006, mouschovias:2006, krumholz:2007}.

At some level and over some spatial scale, an age spread must exist. If we look at a region containing many star forming ``events" taking place over a large volume, then we expect some dispersion in their star forming histories (see section~\ref{sec:kinematics}). How can we measure the age spreads and what are the appropriate sizes to consider?

The co-existence of class I, II and III PMS stars (i.e., in progressively advanced evolutionary stages) in the same cluster/cloud/region is sometimes taken as evidence for an age spread, under the assumption that disk evolution is monotonic. Others attribute this to variation in disk dissipation time scales.  An order of magnitude spread in luminosity at a given temperature is almost ubiquitous in the HRDs of young clusters and star-forming regions with mean age $<10$\,Myr (an example for the Orion Nebula Cluster is shown in Fig.~\ref{fig:oncspread}). The interpretation of this luminosity dispersion has important consequences for age estimation. If it is taken as evidence for an age spread, then that spread typically has FWHM $\simeq 1$\,dex in $\log \tau$, and would suggest a star forming epoch lasting $\geq 10$\,Myr \citep{palla:2000,huff:2006} and would call into question the use of very young clusters as fiducial ``points" to investigate early evolutionary processes.  Others argue that luminosity estimates could be inaccurate due to observational uncertainties or astrophysical nuisance parameters such as binarity, accretion, and variability (see section~\ref{sec:hrd}) or even that the relationship between HRD position and age is scrambled at early ages by variable accretion history or magnetic activity. If so, then an age determined from the HRD for an individual star might be in error by factors of a few, and have an uncertain systematic bias.

%

\subsection{Observed age spreads and observational uncertainties}

The luminosity of a low mass PMS star declines as $\tau^{-2/3}$ at almost constant $T_{\rm eff}$. The typical Gaussian dispersion in $\log L$ (e.g., in Fig.~\ref{fig:oncspread}) is $\simeq 0.25-0.3$\,dex, with a consequent pseudo-Gaussian dispersion of inferred $\log \tau$ of $\simeq 0.4$\,dex. It has been realized for some time \citep*[see discussion in section~\ref{sec:hrd} and, e.g.,][]{hartmann:2001, hillenbrand:2009} that a mixture of uncertainties, both observational and astrophysical 
might contribute significantly to this dispersion. 
Much recent work has focused on attempting to assess and simulate these uncertainties in an effort to establish what fraction of the luminosity dispersion is attributable to a genuine age spread, in some cases concluding that this fraction is large \citep{hillenbrand:2008, slesnick:2008, preibisch:2012}, in other cases quite small \citep{dario:2010b, reggiani:2011}. 

The conclusions from the studies above are critically dependent on an accurate assessment of the observational uncertainties and their (probably non-Gaussian) distribution and also on establishing the bona fide cluster membership of stars contributing to the age distribution. There is often a small (few per cent) tail of very low luminosity objects (see Fig.~\ref{fig:oncspread}) with implied ages a lot older than 10\,Myr. It is possible that some of these objects are not genuine cluster members, but unassociated field stars (A classic, lingering case of this are the sources HBC 353-357 associated for many decades with Taurus, but more likely is that they are members of the background Perseus complex, raising their luminosity). 
Even if they share the mean cluster kinematics it is possible they were captured during cluster formation \citep{pflamm:2007}. Another possibility that seems borne out for some fraction of these ``older" objects is that they are young objects with circumstellar disks that are viewed primarily through scattered light \citep{slesnick:2004, guarcello:2010}. Finally, it should be remembered that what is viewed on the sky could be a projection of two or more separate star forming regions with different ages and this would be misinterpreted as requiring an extended cluster formation timescale \citep[e.g.,][]{alves:2012}. 
The young Galactic super star cluster NGC~3603 is an excellent example of the confusion that can arise. \citet{beccari:2010} show CMDs from the cluster and surrounding area implying age spreads of more than 20\,Myr. However, from a much smaller ``core" region where extinction is uniform and disks may have photoevaporated, and with a proper-motion selected sample, \citet{kudryavtseva:2012} claim that star formation is "instantaneous". This cautions that one must consider the size of the region being observed, ensure that membership of the cluster is secure and carefully assess the observational uncertainties.  There is some contrary evidence from multi-wavelength studies that some star-forming regions show age spreads of $\sim2$ Myr \citep{wang:2011w, bik:2012f}.

\subsection{Luminosity spreads without age spreads?}        
\label{sec:lumspread}

When does age start and what does $\tau = 0$ mean?  The interpretation of a luminosity spread as due to an age spread implicitly relies on the veracity of PMS isochrones in the HRD or at least on the assumption that there is a one-to-one mapping between luminosity and age. In fact the age of a PMS star deduced from the HR diagram will depend on the definition of $\tau=0$ and the initial conditions will remain important to any age estimate until they are erased after a Kelvin-Helmholtz (K-H) timescale. \citet{stahler:1988} has advocated the use of a ``birthline," where PMS deuterium burning defines a core mass-radius relationship. However, it has long been appreciated that early accretion can change this relationship providing a spread in birthlines that is then propagated for the first few Myr of the PMS lifetime \citep{mercer:1984, hartmann:1997, tout:1999, baraffe:2002}.

New observations and theoretical calculations have suggested that accretion onto embedded protostars can be at very high rates ($\dot {M} \geq 10^{-5}M_{\odot}\,$yr$^{-1}$, but interspersed with longer periods of much lower accretion rates \citep[e.g.,][]{vorobyov:2006, enoch:2009, evans:2009, vorobyov:2010}.  Models of early evolution incorporating high levels of accretion during the class I PMS phase have been presented by \citet{baraffe:2009}, \citet{hosokawa:2011}, and \citet{baraffe:2012}. \citet{baraffe:2009} presented models of ``cold accretion", where a negligible fraction, $\alpha \sim 0$, of the accretion kinetic energy is absorbed by the star. In such circumstances, and where the accretion timescale is much shorter than the K-H timescale, the response of the star is to contract and appear in the HRD mimicking a much older (and smaller) star, and where it will remain for a K-H timescale of $\sim 10$\,Myr. 

On the other hand, if a significant fraction ($\alpha \geq 0.2$) of the accretion energy is absorbed, the star swells and becomes more luminous but, because its K-H timescale becomes much smaller, it quickly returns to follow the appropriate non-accreting isochrone.  Variations in $\alpha$ and the accretion rate (or more precisely the amount of material accreted on a short timescale) would scatter a 1\,Myr co-eval population in the HRD between 1 and 10\,Myr non-accreting isochrones. \citet{hosokawa:2011} present similar models, concluding that the effects discussed by \citet{baraffe:2009} apply only to PMS stars seen with $T_{\rm eff}> 3500$\,K once accretion has ceased and that since luminosity spreads are observed at lower temperatures too, then it is possible that inferred age spreads are genuine. \citet{baraffe:2012} reconciled these views to some extent, finding that for stars with a small final mass (and $T_{\rm eff}$), that initial seed protostellar mass was an important parameter, and that at smaller values than assumed by \citet{hosokawa:2011} cold-accretion {\it did} lead to significant post-accretion luminosity spreads in stars with differing accretion histories. Both ages and masses from the HRD would be over-estimated using non-accreting isochrones. Theoretical arguments are hampered by the lack of very detailed 3D numerical simulations of the formation of the protostellar core and the accretion process.

Another possible contributor to the dispersion in the HRD is that magnetic fields or significant spot coverage will inhibit the flow of energy out of the star and may change the luminosity and radius significantly and hence also $T_{\rm eff}$ \citep{spruit:1986, chabrier:2007, feiden:2012}. Such anomalies have been measured in close, magnetically active, eclipsing binary systems, where $T_{\rm eff}$ may be decreased by 10\% or more. It is possible that the effects may be more significant in fully convective PMS stars \citep{jackson:2009, macdonald:2013}, but no definitive investigation of magnetic activity or spottiness versus HRD position has been performed. 


\subsection{Testing age spreads independently of the HRD}       
\label{sec:runaway}

The reality of significant ($\sim 10$\,Myr) age spreads can be tested independently of the HRD using other age indicators. Younger stars ought to have larger radii, lower gravities, lower core temperatures and, for a given angular momentum, slower rotation rates, than older stars of similar $T_{\rm eff}$ in the same cluster.  \citet{jeffries:2007a} modeled the distribution of projected stellar radii (see section~\ref{sec:rsini}) in the ONC, demonstrating that a single age failed to match the broad spread of $R \sin i$ and that a more consistent match was found by assuming the radii were those given by HRD position. This implies a genuine factor of $\sim 3$ spread in radii at a given $T_{\rm eff}$ and that contributions to the luminosity dispersion from observational uncertainties and variability are small. Whilst this does support the notion of a large luminosity dispersion it does not necessarily indicate a wide age spread. 

Evolutionary models featuring efficient convection or large mixing length begin to exhibit surface Li depletion at $\simeq 10$\,Myr at $M \sim 0.5\,M_{\odot}$ (see section~\ref{sec:li}). A number of authors find stars at these or even cooler spectral types that appear to have lost Li despite being members of young clusters/associations like the ONC and Taurus. Some of these stars also show evidence for accretion disks, requiring a careful veiling correction \citep{white:2005, palla:2007}, but in others the weak Li line is unambiguous \citep{sacco:2007}.  The ages estimated from the HRD are generally younger than implied by the level of Li depletion, and the masses lower -- a feature that may persist to older ages \citep{yee:2010}. 

These Li-depleted stars support the notion of a (very) wide age spread, but represent at most a few percent of the populations of their respective clusters. 
Yet this Li-age test may not be independent of the accretion history of the star: \citet{baraffe:2010} suggest that the same early accretion that would lead to smaller luminosities and radii, and apparently older ages compared to non-accreting models, will also lead to higher core temperatures and significant Li depletion. \citet{sergison:2013} found a modest correlation between the strength of Li absorption and isochronal age in the ONC and NGC~2264, but no examples of the severely Li-depleted objects that \citet{baraffe:2010} suggest would be characteristic of the large accretion rates required to significantly alter the position of a PMS star in the HRD.

\citet{littlefair:2011} examined the relationship between rotation and age implied by CMD position within individual clusters. PMS contraction should lead to significant spin-up in older stars, or if the ``disk locking" mechanism is effective \citep[for stars which keep their inner disks for that long;][]{konigl:1991}, rotation periods might remain roughly constant. In all the examined clusters the opposite correlation was found -- the apparently "young" objects were faster rotating than their "older" siblings -- questioning whether position in the CMD truly reflects the stellar age and possibly finding an explanation in the connection between accretion history and present-day rotation.

The presence or not of a disk or accretion acts as a crude age indicator (see section~\ref{sec:disks}). In the scenario where most stars are born with circumstellar disks, and that disk signatures decay monotonically (on average) over time-scales of only a few Myr, any age spread greater than this should lead to clear differences in the age distributions of stars with and without disks. Some observations match this expectation \citep{bertout:2007, fang:2013b}, but many do not and the age distributions of stars with and without accretion signatures cannot be distinguished \citep{dahm:2005, winston:2009, rigliaco:2011}. \citet{jeffries:2011} performed a quantitative analysis of stars in the ONC with ages given by \citet{dario:2010} and circumstellar material diagnosed with Spitzer IR results. The age distributions of stars with and without disks were indistinguishable, suggesting that any age spread must be smaller than the median disk lifetime.

Kinematic expansion is unlikely to constrain any age spread, but there are examples of runaway stars that apparently originated from star forming regions. Even with excellent kinematic data, we would need several runaways from the same region in order to draw any conclusions about age spreads. 

Pulsations among intermediate mass stars in the instability strip (see section~\ref{sec:seismology}) have potential for diagnosing age spreads if good pulsation data for a number of stars in the same cluster can be obtained. Whilst not independent of the evolutionary models, the ages determined in a seismology analysis are not affected by the same observational uncertainties as the HRD and can be considerably more precise \citep[e.g.,][]{zwintz:2013}.

\smallskip
\noindent{\bf Summary on age spreads:}

The issue of whether significant age spreads exist is not settled either observationally or theoretically and the degree of coevality may vary in different star forming environments. It is clear that the age implied by the HRD for an individual star in a young ($\leq 10$\,Myr) cluster should not be taken at face value, especially if that star has circumstellar material. At best there are large observational uncertainties and at worst the inferred ages are meaningless, almost entirely dependent on the accretion history of the star. It is possible that mean isochronal ages for a group of stars holds some validity but could be subject to an uncertain systematic bias.


\section{
THE EFFECTS OF COMPOSITION
}
\label{sec:additional}

There are several additional factors related to the composition of stars that influence age determinations because they affect both observations and models.  In particular, what initial composition should be adopted for young ($<$100 Myr-old) stars in the solar vicinity? Is the present-day solar photospheric or estimated proto-solar abundance mix adequate?

Evolutionary tracks require mass fractions of hydrogen, helium, and
metals -- and of course a choice of individual mass fractions for the
``metal" elements. To first order, the young stars in the solar
vicinity appear to be ``solar" in composition. Unfortunately, it is
still unclear exactly what the solar composition is, as there is
disagreement between the solar metal fraction derived using comparing
solar spectra to new 3D stellar atmosphere models, and that derived
using helioseismological sound speed profiles
\citep{antia:2006,asplund:2009,caffau:2010}.  As we noted above, LDB ages are {\it not} sensitive to changes in the composition of models and thus those ages are robust.

\subsection{
Helium abundances
}

Over the past decade it has become clear from the CMDs of some globulars that apparently similar stars can have a range of He abundances.  He is important in determining the structure and hence evolution of a star, yet it remains stubbornly difficult to measure observationally.  The general assumption is that other stars have the same He as the Sun, or that there is a general trend in Galactic evolution of He increasing with [Fe/H] over time.  It needs to be recognized that uncertainty in He adds fundamental uncertainty to most aspects of PMS stars, including their ages.

Recently there has grown a general consensus 
that the protosolar helium mass fraction was in the range
$Y_{\odot}^{\rm init}$ $\simeq$ 0.27-0.28
\citep{grevesse:1998,asplund:2009,lodders:2009,serenelli:2010}. Few He
abundances measurements are available for nearby young $<$100 Myr-old
stars.  Spectra of young B-type stars in nearby clusters shows that
their He abundances are more-or-less similar to the modern
photosphere, albeit with high dispersion which appears to be due to
rotationally-induced mixing \citep{huang:2006}.  Table 9 of
\citet{nieva:2012} shows a useful modern comparison between abundances of
He, C, N, O, Ne, Mg, Si, and Fe amongst nearby B-stars, other young
samples of stars, the interstellar medium, and the Sun. Save for Ne,
there is fairly good agreement between the inferred abundances for
B-type stars in the solar vicinity and the solar photospheric
abundances from \citet{asplund:2005}.  Solar helium abundances
have also been inferred for members of the Pleiades \citep{southworth:2005, an:2007},
and other nearby very young clusters \citep{mathys:2002, southworth:2004, 
southworth:2007,alecian:2007}. 

\subsection{
Metallicity
}

Metallicities of PMS stars can be determined from observations, but with lower precision and accuracy than for main sequence stars. Complications may arise from the effects of starspots and plage regions and non-LTE effects affective spectral line formation \citep{stauffer:2003,vianaalmeida:2009,bubar:2011}.   The majority of young stellar groups investigated which are near the Sun appear to have solar {\it metallicity}. Metallicities consistent with [Fe/H] $\simeq$ 0.0 have been measured amongst nearby young open clusters \citep{randich:2001,chen:2003}. 
Star-forming regions also appear to have metallicities
consistent with solar [Fe/H] $\simeq$ 0.00 \citep{padgett:1996, dorazi:2009, 
vianaalmeida:2009, dorazi:2011},
or perhaps slightly less at $\simeq -0.08$ \citep{santos:2008}.
The cloud-to-cloud metallicity dispersion appears to be low,  $\sim$0.03 dex among six nearby star-forming regions \citep{santos:2008},
and  $\sim0.06$ dex among eleven nearby young associations \citep{vianaalmeida:2009}. 
There are claims of some young nearby clusters having [Fe/H] upwards of $\sim$0.2 dex \citep{monroe:2010}. 
For 15 open clusters within 1 kpc of the Sun from the compilation of \citet{chen:2003}, the
mean metallicity is [Fe/H] = $-0.02\pm0.04$. The rms dispersion is $\pm$0.16 dex, however this is likely an
upper limit to the real scatter as the metallicities come from many
studies, and the size of the uncertainties is not listed. The results
seem consistent with most nearby young stars having metallicity
near solar, with 1$\sigma$ dispersion $<$0.1 dex. 

We conclude that the input protostellar composition for 
evolutionary models for $<$100 Myr-old stellar populations in the
solar vicinity should probably have $Z/X$ similar to the modern-day
solar photosphere (hence [Fe/H] $\simeq$ 0.0), and
helium mass fraction similar to
protosolar. A good starting point for models of nearby young stellar
populations might be the recommended ``present-day cosmic matter in
the solar neighborhood" abundances from \citet{przybilla:2008} which
are $X$ = 0.715, $Y$ = 0.271, $Z$ = 0.014, $Z/X$ = 0.020.

\section{
CONCLUSIONS AND RECOMMENDATIONS
}
\label{sec:conclusions}

\subsection{
Age techniques across the HRD
}
\label{sec:hrd}

Table 2 categorizes the applicability of the different methods discussed above in various domains of stellar mass and age.  The individual techniques within each cell are listed in order of reliability in the opinion of the present authors.

\begin{table*}
\begin{tabular}{lcccccccc}
\hline
\hline
                &  1-10 Myr                        &    $\sim$10-100 Myr       &  $>$100 Myr              \\
        \hline 
$<$0.1 $M_\odot$  &   isochrones, gravity, $R \sin i$, seismology?  & LDB,  isochrones, gravity     &      LDB,  isochrones     \\       
0.1-0.5 $M_\odot$  &  isochrones, gravity, $R \sin i$, disks          &     isochrones, Li, gravity  &    rotation/activity       \\       
0.5-2.0 $M_\odot$  &    isochrones, disks                            &     Li  &    rotation/activity         \\       
$>$2.0 $M_\odot$  &    isochrones, seismology, R-C gap           &      isochrones, seismology      &    isochrones \\       
\hline
\hline
\end{tabular}
\caption{
Useful age-dating methods for various mass- and age ranges in the H-R diagram
}
\end{table*}

\subsection{
General comments
}

We have summarized the methods known to us by which one might estimate the age of a young star or group.  Two conclusions are noteworthy.

First, we feel that the evidence supports using the age {\it scale} established by the Lithium Depletion Boundary (sec.\,\ref{sec:ldb}) for young clusters with age $>20$\,Myr.  The number of clusters with LDB ages is modest, but they span a broad range of age and the LDB scale avoids the problems and uncertainties that have arisen in computing models for intermediate-mass stars at the main sequence turn-off points of young clusters.  Adopting the LDB ages implies that a moderate amount of convective core overshoot or rotation needs to be included in models of intermediate mass stars, a conclusion now supported by the analysis of oscillations, detected by {\it Kepler}, in stars slightly more massive than the Sun \citep{silva:2013}.

Second, below 20\,Myr there is no well-defined absolute age scale. This should be considered carefully before using, for example, the median disk lifetime and ratios of different protostellar classes to estimate the lifetimes of protostellar phases. Kinematic ages ought to provide a timeline independent of any stellar physics uncertainties but, frustratingly, attempts to estimate the ages of groups of very young stars through their kinematics appear to consistently fail (see sec. \ref{sec:kinematics}).  

What does work?  All the empirical methods (rotation, activity, lithium, IR excesses) either have inherently large scatters at any one age, to a degree that exceeds any age trend, or they are the very properties we hope to study as essential aspects of PMS evolution.  Placement of stars and associations in CMDs and HRDs remains fundamental in this field, despite the many and known difficulties.  PMS isochronal ages can have good precision, allow groups of stars to be ranked, but have model-dependent systematic uncertainties of at least a factor of two below 10\,Myr. Ages from the UMS or MSTO are probably more reliable, but often less precise because there are fewer high mass stars.

As we asked at the beginning, can we, in fact, establish reliable and consistent ages for young stars that are independent of the phenomena being studied?  Our answer is probably "yes" for ages $>20$\,Myr, but has to be ``no'' for younger objects, tempered by some progress, and with expectations of that progress continuing.  The biggest improvement, by far, that we can anticipate is Gaia.  All of stellar astrophysics has great expectations for the success of that mission, and it can seem as though all hope is vested in it.  Nevertheless, the expectations are driven by realistic estimates of Gaia's performance.  For PMS studies, Gaia will make it possible to measure directly the precise distances to individual stars in star-forming regions out to Orion and beyond.  That obviates uncertainties due to extinction and reddening and allows the full 3D structure of those regions to be seen.  Gaia should detect more and lower-mass runaway stars and its precise astrometry, when combined with good radial velocities, should reveal precise 3D velocities that could lead to accurate kinematic ages.

Related to Gaia, the Gaia-ESO spectroscopic survey \citep{gilmore:2012} will provide large, homogeneous data sets in $\sim 30$ young clusters, yielding uniformly determined abundances (Fe, Li and others) along with RVs more precise than possible with Gaia ($<0.5$~km~s$^{-1}$), with which to determine membership.

\subsubsubsection{Recommendations, and a final thought}
\noindent
{\bf To theorists and modelers:} Please include the pre-main sequence in your published evolutionary tracks (most now do), please produce very dense grids of evolutionary tracks and isochrones by mass and age, and please carefully consider assumptions about composition. 

\noindent
{\bf To observers:} When estimating ages of stars using evolutionary tracks, we recommend that you also try your technique on a few test stars to make sure that your results make sense.  As a case in point, old field M dwarfs can have luminosities and effective temperatures which make them appear to be PMS and $<$100 Myr. This is because evolutionary tracks have difficulty predicting the main sequence among the low-mass stars. Age-dating groups of stars (clusters, associations, multiple stars) will most likely always be more accurate than age-dating individual stars (save perhaps special cases using asteroseismology). 

\noindent
{\bf An idea:}
There should be a deuterium equivalent of the LDB and because deuterium burns at lower temperatures, it occurs earlier and in lower mass objects. Stars of mass $0.1\,M_\odot$ deplete their D in just 2\,Myr, but it takes 20\,Myr to reach D depletion in a $0.02\,M_\odot$ brown dwarf \citep{chabrier:2000}. Hence for a coeval group of stars and brown dwarfs with an age in this interval there should be a luminosity below which D is present, but above which it is absent. Detecting and measuring D in stars is challenging, but M-type spectra have molecular bands (perhaps HDO, CrD) that might be amenable to this, and the goal is not a precise measurement of D/H but instead just detection of the isotope. Potentially then, this could yield absolute ages below 20\,Myr, although the definition of $t=0$, the possibility of age spreads, the role of initial conditions and the early or ongoing accretion (of D-rich material) would certainly make interpretation challenging.


\smallskip
\noindent
\textbf{Acknowledgments} 

EEM acknowledges support from NSF grant AST-1008908.  The careful reading by the referee was appreciated.

\bigskip

\bibliographystyle{ppvi_lim1.bst}
\bibliography{cit}

\end{document}